\documentclass[twocolumn]{aastex63}

\usepackage{verbatim}

\newcommand{\lya}{Ly$\alpha$}

\newcommand{\heii}{He\,{\sc ii}}
\newcommand{\ciii}{C~{\sc iii}]}
\newcommand{\oiii}{[O~{\sc iii}]}
\newcommand{\oii}{[O~{\sc ii}]}
\newcommand{\oiiiI}{O~{\sc iii}]}
\newcommand{\nv}{N\,{\sc v}}

\newcommand{\heiil}{\heii~$\lambda$1640}
\newcommand{\ciiil}{\ciii~$\lambda$1909}

\newcommand{\nvli}{\nv~$\lambda$1239}
\newcommand{\nvlii}{\nv~$\lambda$1243}
\newcommand{\nvld}{\nv~$\lambda\lambda$1239,1243}
\newcommand{\oiiil}{\oiii~$\lambda\lambda$4959,5007}

\newcommand{\oiiiIl}{\oiiiI~$\lambda$1665}
\newcommand{\oiild}{\oii~$\lambda\lambda$3726,3729}

\newcommand{\civ}{C\,{\sc iv}}
\newcommand{\civl}{\civ~$\lambda\lambda$1548,1551}

\received{}
\revised{}
\accepted{}
\submitjournal{ApJ}

\shorttitle{LAEs at $z\approx$3.1}
\shortauthors{Guo et al.}

\graphicspath{{./}{figures/}}

\begin{document}

\title{A Spectroscopic Survey of Ly$\alpha$ Emitters at $z\approx3.1$ over $\sim$1.2 Deg$^2$}

\author{Yucheng Guo}
\affiliation{Kavli Institute for Astronomy and Astrophysics, Peking University, Beijing 100871, China}
\affiliation{Department of Astronomy, School of Physics, Peking University, Beijing 100871, China}

\author[0000-0003-4176-6486]{Linhua Jiang}
\altaffiliation{jiangKIAA@pku.edu.cn}
\affiliation{Kavli Institute for Astronomy and Astrophysics, Peking University, Beijing 100871, China}
\affiliation{Department of Astronomy, School of Physics, Peking University, Beijing 100871, China}

\author{Eiichi Egami }
\affiliation{Steward Observatory, University of Arizona, 933 N. Cherry Avenue, Tucson, AZ 85721, USA}

\author{Yuanhang Ning}
\affiliation{Kavli Institute for Astronomy and Astrophysics, Peking University, Beijing 100871, China}
\affiliation{Department of Astronomy, School of Physics, Peking University, Beijing 100871, China}

\author{Zhen-Ya Zheng}
\affiliation{CAS Key Laboratory for Research in Galaxies and Cosmology, Shanghai Astronomical Observatory, Shanghai 200030, China}

\author{Luis C. Ho}
\affiliation{Kavli Institute for Astronomy and Astrophysics, Peking University, Beijing 100871, China}
\affiliation{Department of Astronomy, School of Physics, Peking University, Beijing 100871, China}

\begin{abstract}

We present a spectroscopic survey of Ly$\alpha$ emitters (LAEs) at $z\approx3.1$ in the Subaru {\it XMM-Newton} Deep Survey Field. This field has deep imaging data in a series of broad and narrow bands, including two adjacent narrow bands NB497 and NB503 that have allowed us to efficiently select LAE candidates at $z\approx3.1$. Using spectroscopic observations on MMT Hectospec and Magellan M2FS, we obtained a sample of 166 LAEs at $z\approx3.1$ over an effective area of $\sim$1.2 deg$^2$, including 16 previously known LAEs. This is so far the largest (spectroscopically confirmed) sample of LAEs at this redshift. We make  use of the secure redshifts and multi-band data to measure spectral properties such as \lya\ luminosity and rest-frame UV slope. We derive a robust \lya\ luminosity function (LF) that spans a luminosity range from $\sim10^{42.0}$ to $>10^{43.5}$ erg s$^{-1}$. Significant overdense and underdense regions are detected in our sample, but the area coverage is wide enough to largely suppress the effect from such cosmic variance. Our \lya\ LF is generally consistent with those from previous studies at $z \sim 3.1$. At the brightest end of the LF, there is a tentative detection of a density excess that is not well described by the Schechter function. The comparison with the LFs at other redshifts suggests that the \lya\ LF does not show significant evolution at $2<z<5$. Finally, we build the composite spectra of the LAEs and detect the \nvli\ and \civl\ doublet emission lines at significance of $\sim 4 \sigma$, suggesting very hard radiation fields in (some of) these LAEs.

\end{abstract}
\keywords{High-redshift galaxies(734); Lyman-alpha galaxies(978); Galaxy properties(615)}

\section{Introduction} \label{sec:intro}

\begin{figure*}[ht!]
\plotone{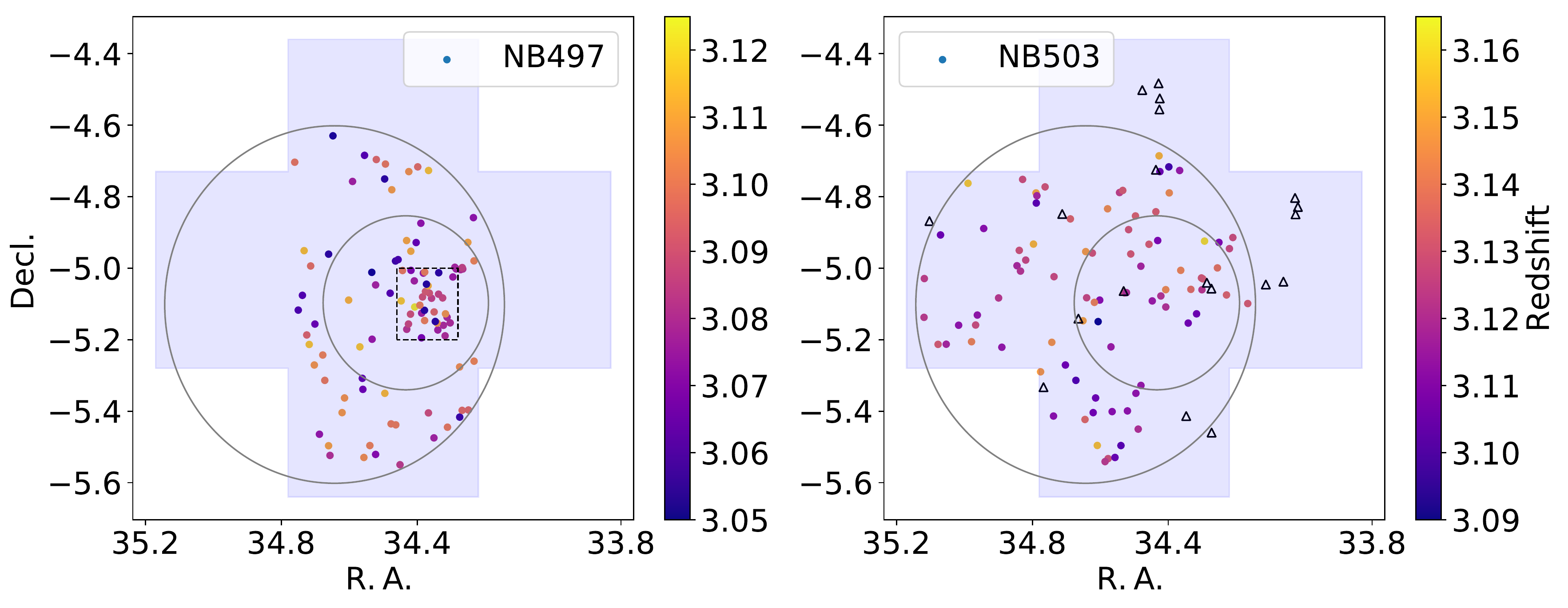}
\caption{Demonstration of our spectroscopic observations of $z \approx 3.1$ LAEs in the SXDS field. The grey area indicates the SXDS field, or the coverage of the Subaru imaging data. The large and small circles denote the pointings of the Hectospec and M2FS observations, respectively. The color-coded dots represent our spectroscopically confirmed LAEs. The LAEs detected in NB497 and NB503 are shown in the left and right panels, respectively. The triangles in the right panel represent the LAEs identified by \citet{ouchi08}. The dotted rectangle in the left panel represents an overdense region.   \label{fig:SXDS}}
\end{figure*}

In the past decades, we have witnessed a significant progress in detecting and studying galaxies at high redshift  \citep[e.g.,][]{madau14,stark16}. There are two common methods to select high-redshift galaxies, the Lyman break technique (the dropout or broadband technique) and the narrowband technique. The narrowband technique combines narrowband and broadband photometry, and searches for strong UV/optical emission lines from star-forming galaxies, such as Ly$\alpha$ and H$\alpha$ emission lines. It usually ensures that the selected galaxy candidates are in small redshift ranges with $\delta z/(1+z)\sim 1\% - 2\%$.

Star-forming galaxies and AGN often produce strong Ly$\alpha$ emission lines. This line is intrinsically the strongest emission line in the rest-frame UV/optical spectrum \citep[e.g.,][]{partridge67,santos16}. In the local universe, a small fraction of galaxies have strong Ly$\alpha$ emission lines because of the low escaping rate of Ly$\alpha$ photons \citep{ciardullo12}. A small amount of dust and/or neutral gas can effectively prevent Ly$\alpha$ photons from escaping from galaxies. However, the Ly$\alpha$ emission line is commonly seen in star-forming galaxies at high redshift \citep[e.g.,][]{shapley03,ciardullo12}. Therefore, the narrowband technique or surveys have been quite successful in searching for high-redshift Ly$\alpha$ emitting galaxies, or Ly$\alpha$ emitters (LAEs).

In recent years, wide-field narrowband surveys have detected a large number of LAEs from $z \approx 2$ to the epoch of reionisation \citep[e.g.,][]{ouchi08,kashikawa11,shibuya12,konno14,matthee14,zheng16,hao18,jiang18,hu19}. Several studies have provided LAE samples at $z \approx 3.1$ \citep[e.g.,][]{gronwall07,ouchi08,ciardullo12,yamada12a,yamada12b,zheng16,sobral18}. Despite the progress that has been made so far, the majority of the galaxies in these samples are photometrically selected candidates. For example, \citet{ouchi08} and \citet{zheng16} spectroscopically observed a fraction of the LAEs in their samples.
\citet{yamada12a} conducted a photometric survey of LAEs at $z \approx 3.1$, and \citet{yamada12b} spectroscopically confirmed 91 LAEs from the photometric sample. This was the largest sample of spectroscopically confirmed LAEs at this redshift.
Overall, there are hundreds of narrowband selected LAE candidates at $z \approx 3.1$, and only a few tens of them have been spectroscopically confirmed. The relatively small number of the confirmed LAEs makes it difficult to compare different results in the literature. There exist large discrepancies in the measurements of the Ly$\alpha$ luminosity function (LF) at $z \approx 3.1$ (and at other redshifts as well). Therefore, a large sample of spectroscopically confirmed LAEs at this redshift is needed.

In this paper, we present our spectroscopic survey of a large sample of LAEs at $z \approx 3.1$ in the Subaru XMM-Newton Deep Survey (SXDS) field. The targets were selected using the deep imaging data taken by the Subaru Suprime-Cam, and the spectroscopic observations were carried out by the Magellan M2FS and MMT Hectospec. We obtained a sample of 166 LAEs over an effective area of $\sim$1.2 deg$^2$ when 16 previously confirmed LAEs are included. We introduce this sample and derive the Ly$\alpha$ LF in this paper. We will measure the physical properties of these LAEs in an upcoming paper.

The layout of this paper is as follows. In Section~\ref{sec:obs}, we introduce our target selection. In Section~\ref{sec:spec}, we present our Magellan M2FS and MMT Hectospec observations. The Ly$\alpha$ and rest-frame UV continuum properties of this sample are provided in Section~\ref{sec:lya_UV}. In Section~\ref{sec:lf} we estimate the sample completeness and derive the Ly$\alpha$ LF. We discuss our results in Section~\ref{sec:discussion} and summarize the paper in Section~\ref{sec:conclusions}. Throughout this paper, all magnitudes are in the AB system. We adopt a $\Lambda-$dominated flat cosmology with $\mathrm{H_0 = 70\, km\, s^{-1}\, Mpc^{-1}}$, $\mathrm{\Omega_m=0.3}$ and $\mathrm{\Omega_\Lambda=0.7}$.

\section{Imaging data and target selection} \label{sec:obs}

In this section, we describe the deep field that we used for our program, the imaging data, and the LAE candidate selection.

\subsection{The SXDS Field} \label{subsec:obs-sxds}

The SXDS field ($\mathrm{02^{h}18^{m}00.0^{s}-05^{\circ}00^{'}00.00^{''}}$; Figure~\ref{fig:SXDS}) covers an area of $\sim$1.2 deg$^2$ \citep{furusawa08}. It has very deep imaging data in a series of broad and narrow bands taken by the Subaru Suprime-Cam.  The SXDS field consists of five subfields, SXDS-C, N, S, E, and W, corresponding to the five pointings of the Suprime-Cam imaging observations.

The SXDS data have been used to search for galaxies at redshift ranging from 2 to 7. For example, \citet{ouchi08} presented a large sample of photometrically selected LAEs at $z \approx  3.1$, 3.7, and 5.7. Some of them were spectroscopically observed. \citet{ouchi10} presented a photometric sample of LAEs at $z \approx 6.5$, and spectroscopically identified 19 LAEs. \citet{konno14} carried out a deep narrowband imaging survey of LAEs at $z \approx 7.3$, and found three LAEs. \citet{matthee15} reported a small sample of bright, photometrically selected LAEs at $z \approx 6.5$. \citet{konno16} presented a large,  photometric sample of 3137 LAEs at $z \approx 2.2$ in five fields including SXDS. \citet{jiang17} performed a spectroscopic survey of LAEs at $z \approx 5.7$ and 6.5 over nearly three square degrees, including SXDS. \citet{chanchaiworawit17} identified 45 LAE candidates around two close, massive LAEs at $z \approx 6.5$ in SXDS. \citet{ota17} detected 20 $z \approx 7.0$ LAE candidates in the Subaru Deep Field and SXDS. \citet{itoh18} presented a large sample of 34 LAE candidates at $z\sim$7.0 in the COSMOS and SXDS fields.

\begin{figure}[t]
\plotone{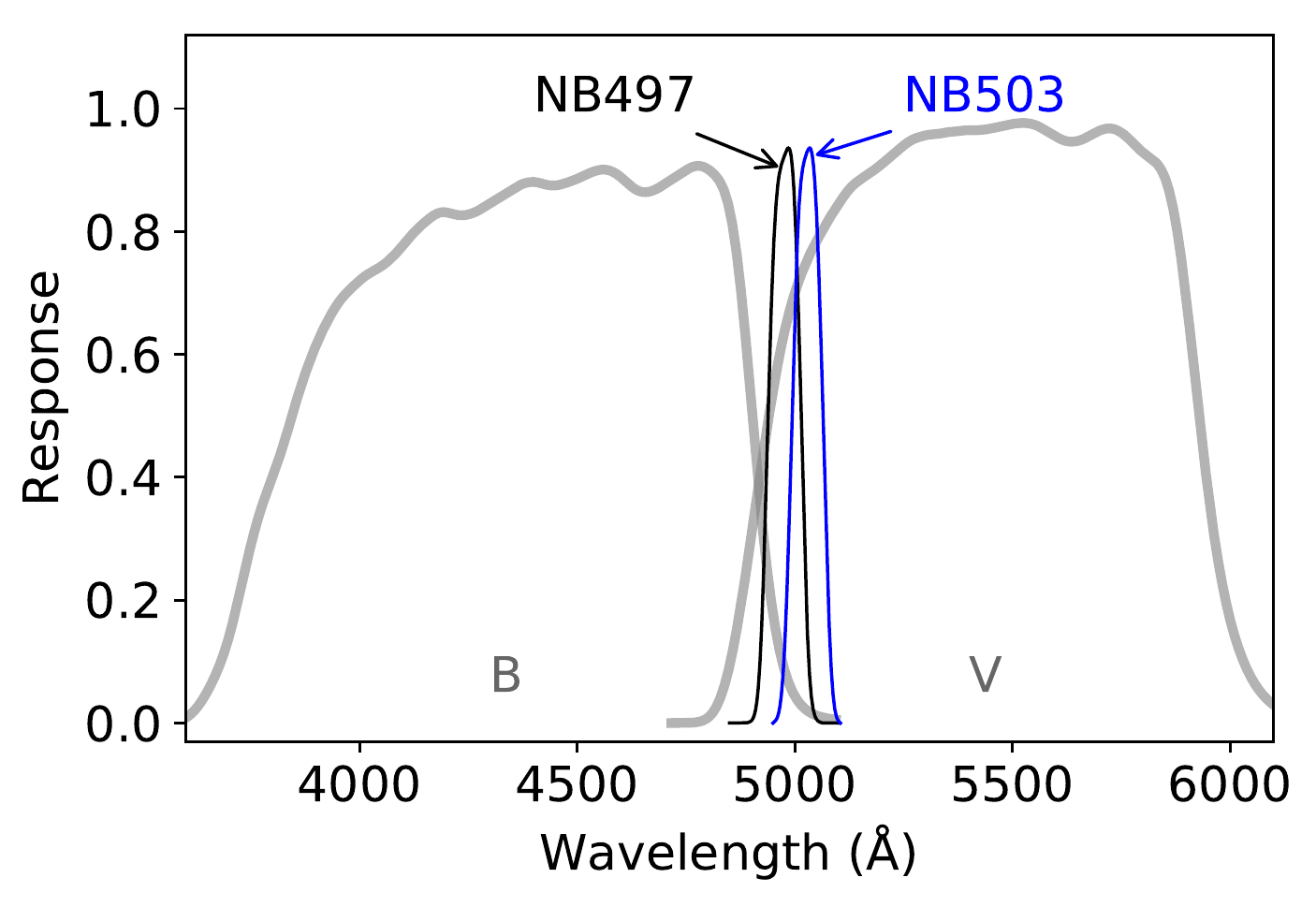}
\caption{The transmission curves of the Subaru Suprime-Cam filters that we have used for our target selection, including two narrowband filters, NB497 and NB503, and two broadband filters $B$ and $V$. \label{fig:filters}}
\end{figure}

We retrieved the raw data of Suprime-Cam in the SXDS field from the archival server SMOKA \citep{baba02}. The images were processed using the Suprime-Cam Deep Field REDuction package \citep[SDFRED;][]{yagi02} and IDL routines by \citet{jiang13}. The details of the image reduction, re-sampling, co-addition, and calibration are provided in \citet{jiang17}.
The depths of the final combined images in five broad bands $B$, $V$, $Rc$, $i'$, $z'$ are 27.9, 27.6, 27.4, 27.4 and 26.2 mag (5$\sigma$ in a $2''$ diameter aperture), respectively. We used two narrow bands NB497 and NB503 to select LAEs at $z \approx 3.1$. NB497 has a central wavelength of $\sim4986$ \AA\  with a full width half maximum (FWHM) of $78$ \AA. NB503 has a  central wavelength of $\sim5030$ \AA\ with a FWHM of $74$ \AA. The transmission curves of the two filters are shown in Figure~\ref{fig:filters}. These two narrowbands correspond to the detection of LAEs at $z\approx3.1$. The NB503-band images cover all five subfields, and the NB497-band images cover the SXDS-C, N, S fields only. The depth in NB497 reaches 25.9 mag, and the NB503 image is about 0.9 mag shallower. The photometric depths vary slightly in five different subfields ($\pm 0.1$ mag).

\citet{ouchi08} and \citet{yamada12a} used the same sets of raw images as we did, and their final images in NB497 and NB503 appear to be slightly deeper than ours. The reason is as follows. The pipeline SDFRED that they used smooths images to match the same point spread function (PSF), usually the worst PSF in the images, before it combines processed individual images. For example, \citet{yamada12a} smoothed their images to $1\arcsec$, while the best PSF was $0\farcs65$. This image smoothing suppresses the background fluctuation and introduces correlated noise within nearby pixels. A direct consequence is that aperture photometry would underestimate background noise, and thus overestimate image depth. We did not smooth images. Instead, we used PSF as a weight and obtained better PSFs in the combined images. The PSFs in our three NB497-band images are $0\farcs65$, $0\farcs81$, and $0\farcs72$. The PSFs in our five NB503-band images are $0\farcs67$, $0\farcs88$, $0\farcs83$, $0\farcs63$, and $0\farcs61$.

\subsection{Candidate Selection}\label{subsec:obs-selection}

We selected LAE candidates at $z\approx3.1$ using the images in the $B$, $V$, NB497, and NB503 bands (Figure~\ref{fig:filters}). The PSF sizes in $B$ and $V$ are better than those in NB497 and NB503. We smoothed the $B$- and $V$-band images so that their PSF FWHMs match the PSF size of the NB497- or NB503-band image. Object detections were performed on the narrowband images. Broadband photometry was done on the narrowband-detected objects using the dual image mode by SExtractor \citep{bertin96}. The aperture size was $2''$ in diameter and aperture corrections were applied to obtain the total magnitudes.

\begin{figure}[t]
\plotone{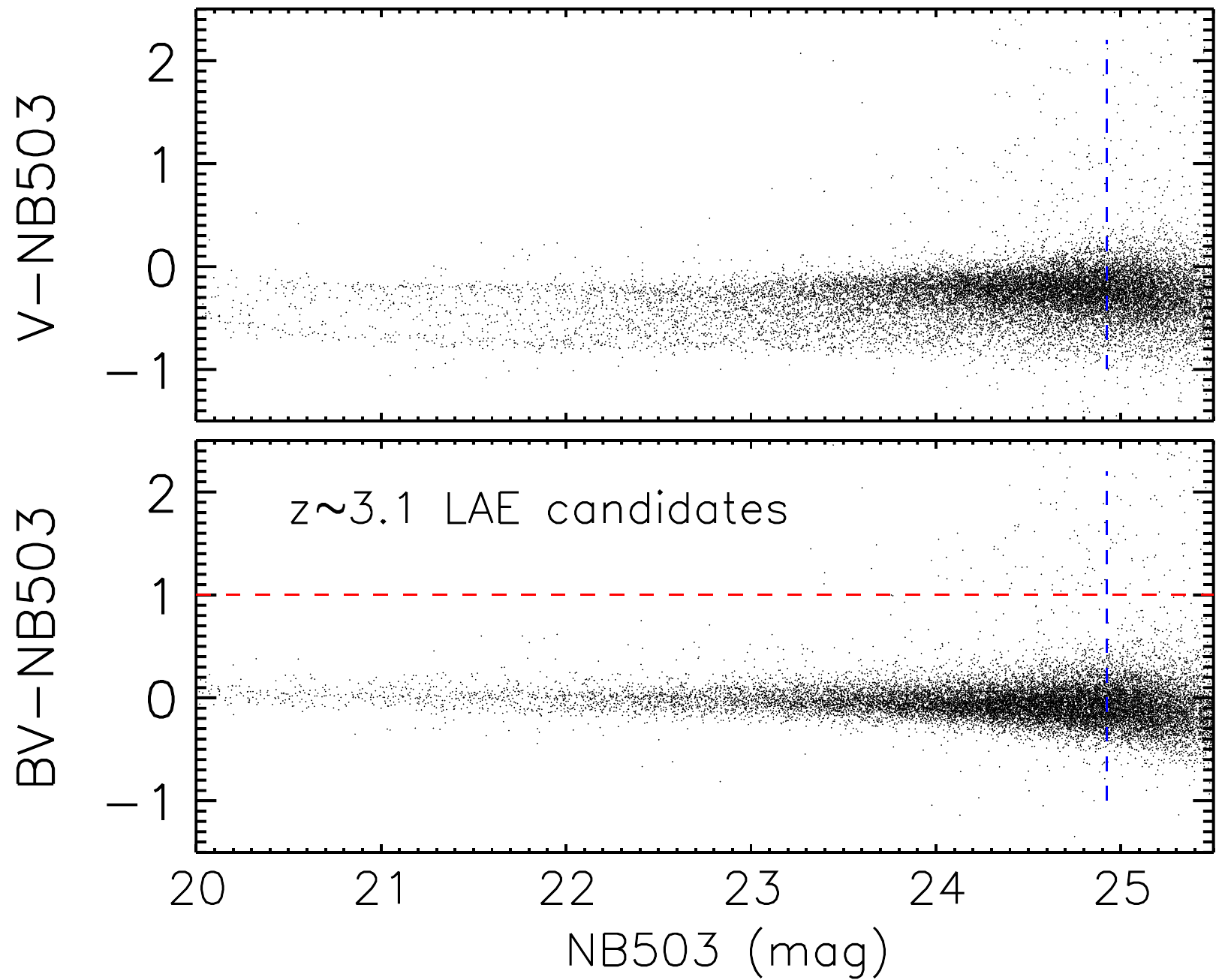}
\caption{Color-magnitude diagrams of the $z\approx3.1$ LAE candidates. We use NB503 as an example. The black dots represent the detected objects in the NB503 band, and the blue dashed lines indicate the 10$\sigma$ detection limits. In the lower panel, the red dashed line indicates our color selection criterion. The upper panel shows the $V$ band magnitude instead of the composite $BV$ magnitude. It is clear that $BV$ performs better than $V$ for our target selection.  \label{fig:targetsel}}
\end{figure}

The candidate selection was mainly done by the selection criterion of $BV - {\rm NB} > 1$ mag, where NB is the narrowband magnitude in NB497 or NB503, and $BV$ is the composite magnitude determined by the $B$-band flux $f_B$ and the $V$-band flux $f_V$ using $f_{BV}=(2f_{B}+f_{V})/3$. We applied this color cut to all detections at $>10\sigma$ in NB497 and all detections at $>9\sigma$ in NB503. We demonstrate our selection in the lower panel of Figure~\ref{fig:targetsel}. For comparison, we show the color-magnitude diagram of NB503 vs. $V$--NB503 for the same objects in the upper panel. We can see that the composite magnitude $BV$ performs better than $V$.
Our color cut criterion is very similar to those used in the literature \citep[e.g.,][]{ouchi08,yamada12a,sobral18}. This criterion roughy corresponds to a \lya\  rest-frame equivalent width (EW) limit of $\sim45$ \AA.

In order to make use of the large number of fibers on Hectospec and M2FS, we included a small amount of weaker LAE candidates whose detections in the narrow bands were slightly below the significance values given above. We also included a small number of candidates with $0.9 < BV - {\rm NB} < 1$ mag. In addition to the $z\approx3.1$ LAEs, we included ancillary targets for spare fibers. We will not present these objects in the paper. These targets had low priorities in our fiber assignment. We visually inspected all candidates and removed sources that were contaminated by bright nearby stars or located in image edges where image quality is significantly lower.

\section{Spectroscopic observations and data reduction} \label{sec:spec}

After we obtained the sample of the LAE candidates, we carried out spectroscopic observations using two multi-fiber spectrographs MMT Hectospec and Magellan M2FS. These spectrographs have large numbers of fibers over large fields-of-view (FoVs). Therefore, they are efficient to observe many targets over a large field. In this section, we describe the spectroscopic observations, data reduction, and target identification. The observations are summarized in Table 1.
The effective area coverage is 0.495 deg$^2$ in NB497 and 0.701 deg$^2$ in NB503.

\begin{deluxetable}{cccccc}[t]
\tablecaption{Summary of Spectroscopic Observations \label{Tab:spec}}
\tablecolumns{6}
\tablewidth{0pt}
\tablehead{
\colhead{No.} & \colhead{Date} & \colhead{Facility} & \colhead{Exp. Time} & \colhead{No. Candidates\tablenotemark{a}} \\
\colhead{(1)} & \colhead{(2)} & \colhead{(3)} & \colhead{(4)} & \colhead{(5)}
}
\startdata
1   & 2016 Oct 06 & Hectospec & 1.61 hrs  & 191            \\
2   & 2016 Oct 09 & Hectospec & 2.00 hrs  & 191            \\
3   & 2017 Sep 28 & Hectospec & 1.18 hrs  & 54             \\
4   & 2017 Sep 28 & Hectospec & 2.00 hrs  & 71             \\
5   & 2017 Oct 01 & Hectospec & 1.50 hrs  & 69             \\
6   & 2016 Nov 28 & M2FS      & 3.00 hrs  & 124            \\
\enddata
\tablenotetext{a}{Number of LAE candidates observed.}
\end{deluxetable}

\subsection{MMT Hectospec Observations}
\label{subsec:spec-Hec}

We observed 191 LAE candidates using Hectospec mounted on the 6.5m MMT. Hectospec provides 300 fibers over a circular FoV of $1^{\circ}$ in diameter. The fiber diameter is ${1.5}''$, with adjacent fibers spaced as closely as $20''$. The pointing was centered at $\mathrm{ R.A.=+02^{h}18^{m}34.35^{s},\; Decl.=-05^{\circ}06^{'}06.00^{''}}$, shown as the large circle in Figure~\ref{fig:SXDS}. We used the 600 gpm grating that provided a resolution of $\sim2$ \AA. This resolution can resolve the [O II] $\lambda\lambda$3726,3729 doublet, a possible contaminant emission line for Ly$\alpha$. The wavelength coverage of the spectra was 4050\AA$-$6500\AA. Several tens of fibers were assigned to blank sky areas for sky subtraction.

To maximize the overall efficiency, we used an observing strategy that fainter LAE candidates received longer exposure time. Meanwhile, we ensured that the exposure time for any candidate was long enough to identify its \lya\ emission line if it is a real LAE at $z\approx3.1$.
The first observation was carried out in October 2016. We then reduced the data from this observation and identified a sample of bright LAEs. Later, we observed other LAE candidates in three observations in September and October 2017. Therefore, the exposure time for individual targets are different. The longest exposure time was 8.3 hrs, and shortest time was 3.6 hrs.

\begin{figure*}[t]
\plotone{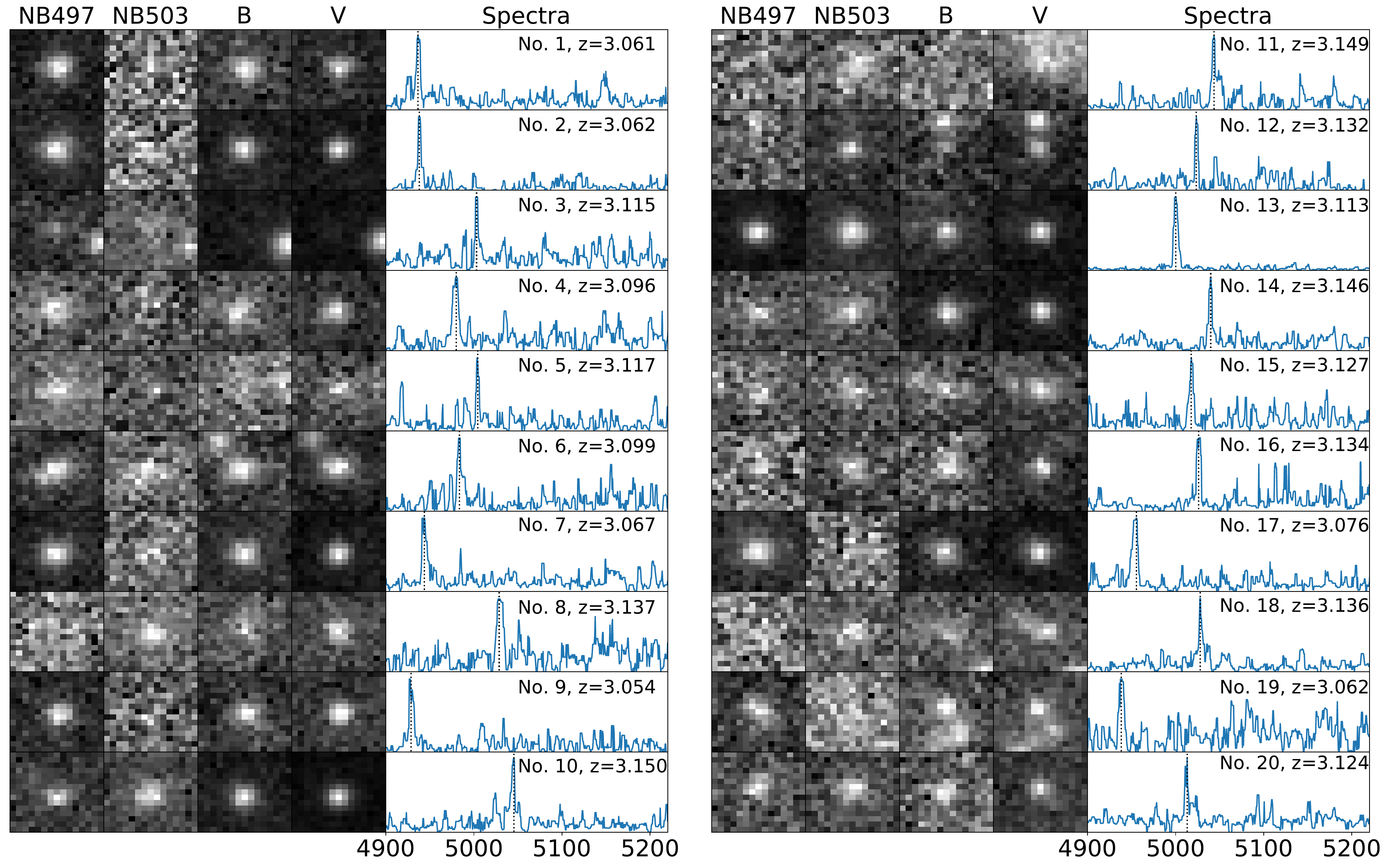}
\caption{Snapshots and spectra of the first 20 LAEs in our sample. For each LAE, we show its stamp images in NB497, NB503, $B$, and $V$, and its spectrum around the Ly$\alpha$ emission line. The vertical dotted line in the spectrum represents the peak of the emission line. The wavelength (the $x$ axis) of the spectrum is in unit of \AA. The information for the whole sample is provided in the electronic version.
\label{fig:spec1}}
\end{figure*}

Our Hectospec data were reduced with the HSRED \footnote{\url{https://www.mmto.org/node/536/}}.  The raw images were de-biased and flat-fielded, and cosmic rays were rejected. Then individual spectra were extracted. Sky templates were produced by averaging the spectra of ``sky fibers", and sky emission was subtracted by scaling the sky templates to match individual science spectra. Wavelength calibration was done by cross-correlating observed spectra against the calibration arc spectra. The final products are one-dimensional (1D), sky-subtracted, wavelength-calibrated,  variance-weighted spectra.

\subsection{Magellan M2FS Observations}

We identified an overdense region of the LAE candidates in the NB497 band. This overdense region is denoted by the dotted rectangle in the left panel of Figure~\ref{fig:SXDS}. We observed 124 LAE candidates in this region in November 2016 using Magellan M2FS. M2FS provides 256 fibers over a circular FoV of $30'$ in diameter. The M2FS pointing was centered at $\mathrm{ R.A.=+34^{h}26^{m}00.45^{s}}$, $\mathrm{Decl.=-05^{\circ}05^{'}48.80^{''}}$, shown as the small circle in Figure~\ref{fig:SXDS}. The total exposure time was 3.0 hrs. The resolving power of the spectra was about 2000.

A standard IRAF routine was used to process the M2FS images. We used the package CCDPROC to correct overscan, subtract bias, and remove dark current. We then used the package HYDRA for the next step. We identified apertures and fit the aperture traces based on the quartz flat frames, and then extracted 1D spectra for quartz, twilight, ThAr arc, and science images. Wavelength calibration was done with the ThAr arc spectra by IRAF tasks IDENTIFY, REIDENTIFY, and REFSPECTRA. A sky spectrum model was derived from ``sky fibers". Sky background was subtracted by scaling the sky spectrum model to match individual source spectra. In the end, we obtained a sky subtracted, wavelength calibrated, 1D spectra for each frame. The final spectrum of each object is the combination (weighted average) of all its individual spectra.

In each Hectospec or M2FS observation, we included about 50 sky fibers, depending on the availability of spare fibers. In addition, we included 5$-$10 relatively bright point targets. They were used as reference stars to check image quality and depth. We excluded some LAE candidates that have already been spectroscopically observed previously. For example, \citet{ouchi08} confirmed 41 LAEs in NB503. We did not observe most of these LAEs, but we will include some of them that were covered by our selection when we calculate \lya\ LF later.

\subsection{LAE Identification}

\begin{figure}[t]
\plotone{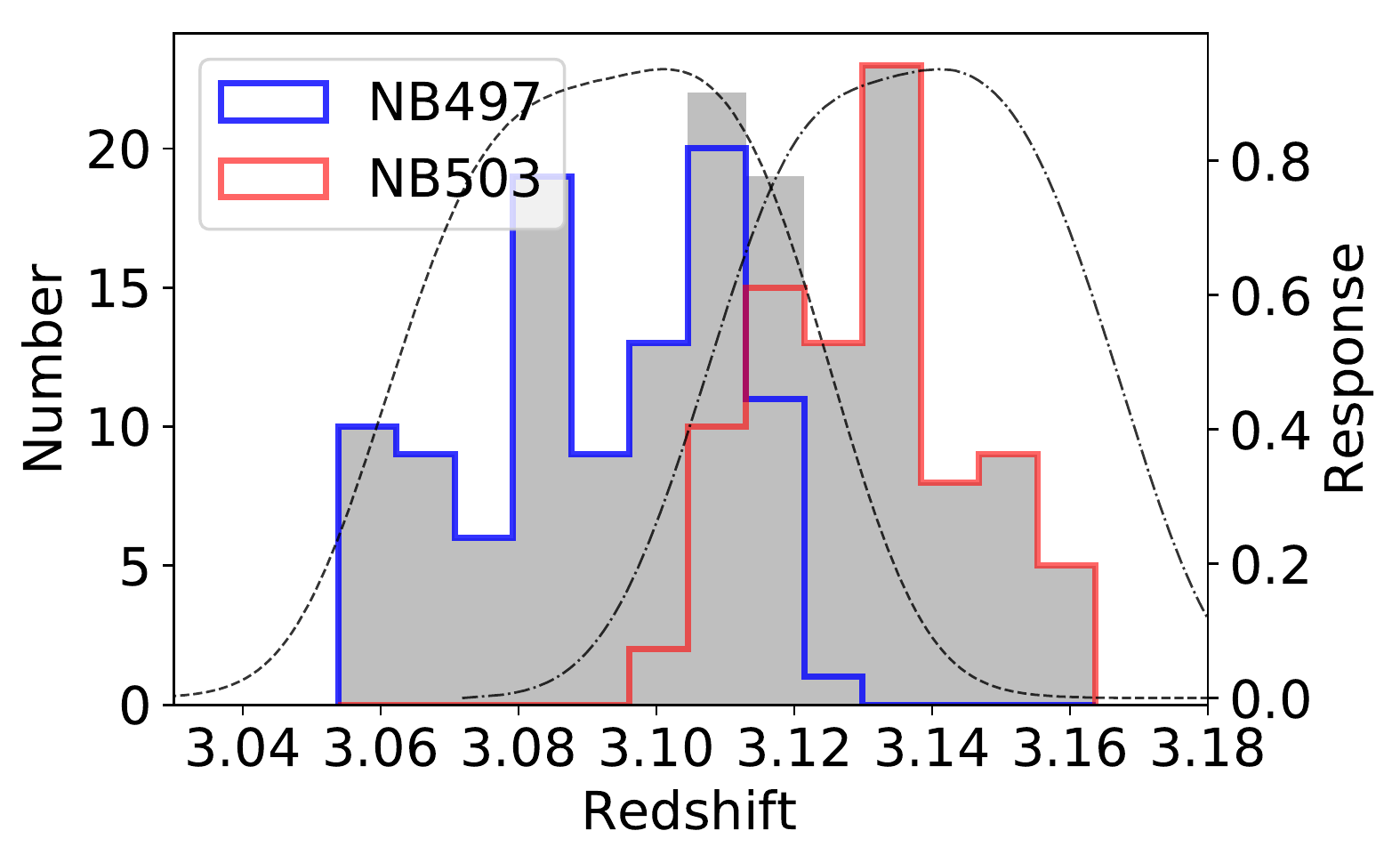}
\caption{The redshift distribution of the LAEs in our sample (shaded histograms). The LAEs detected in NB497 and NB503 are plotted by the blue and red histograms, respectively. The dashed profiles indicate the response curves of NB497 and NB503. Note that some LAEs at $z\sim 3.11$ are detected in both filters. \label{fig:redshift}}
\end{figure}

We identified LAEs based on the 1D spectra (Figure~\ref{fig:spec1}). We searched for line emission features in the expected wavelength range of each LAE candidate. For each identified emission line, we estimated its signal-to-noise ratio (S/N) by stacking pixels around the peak within a window of 7 pixel. A line with S/N $>5$ was treated as a real line detection. Our target selection criteria usually ensure that a detected emission line is a Ly$\alpha$ line. We removed lower-redshift interlopers by checking the whole spectra. For LAEs at $z=3.1$, the possible interlopers are \oiild, H$\beta$ and \oiiil\ emitters. The wavelength coverage of the spectra is large enough to cover all H$\beta$ and \oiiil\ lines if the detected line is one of the three lines. In addition, the spectral resolution is high enough to resolve the \oiild\ doublet, as we mentioned earlier.

We confirmed a total of 150 LAEs at $z \approx 3.1$ from our spectra. In addition, there were 16 LAE candidates in our sample that were not observed by our spectroscopic observations, because \citet{ouchi08} had already confirmed them (mentioned in Section 3.2). When we include these 16 LAEs, our LAE sample consists of 166 LAEs. This sample is the largest, spectroscopically confirmed  sample of LAEs at this redshift. In Figure~\ref{fig:spec1}, we show the spectra of the first 20 LAEs and their snapshots of broadband and narrowband images. The properties of the first 20 LAEs are listed in Table~\ref{tab:sample}. The spectra and properties of the whole sample are provided as the online material.

\begin{deluxetable*}{ccccccccccc}
\tablecaption{The LAE sample \label{tab:sample} }
\tablehead{
\colhead{ID} & \colhead{R.A.} & \colhead{Decl.} & \colhead{Redshift\tablenotemark{a}} &
\colhead{$\mathrm{L_{Ly\alpha}}$} & \colhead{NB497} & \colhead{NB503} & \colhead{$B$} & \colhead{$V$} \\
\colhead{} &  \colhead{(J2000)} & \colhead{(J2000)} &
\colhead{} & \colhead{$\mathrm{(10^{42}\,erg\,s^{-1})}$} & \colhead{(mag)} & \colhead{(mag)} &\colhead{(mag)} & \colhead{(mag)}
}
\startdata
1 & 2 19 00.01 & --5 07 02.2 & 3.061 & 8.73 & $\mathrm{24.51 \pm 0.04 }$ & -- & $\mathrm{26.79 \pm 0.10 }$ & $\mathrm{26.36 \pm 0.08 }$ \\
2 & 2 18 57.07 & --5 04 33.4 & 3.062 & 8.49 & $\mathrm{24.31 \pm 0.03 }$ & -- & $\mathrm{26.00 \pm 0.04 }$ & $\mathrm{25.58 \pm 0.04 }$ \\
3 & 2 18 55.87 & --4 57 03.4 & 3.115 & 2.78 & $\mathrm{25.62 \pm 0.11 }$ & -- & $\mathrm{27.24 \pm 0.15 }$ & $\mathrm{26.90 \pm 0.13 }$ \\
4 & 2 18 53.99 & --5 11 12.6 & 3.096 & 2.04 & $\mathrm{25.20 \pm 0.08 }$ & -- & $\mathrm{26.74 \pm 0.10 }$ & $\mathrm{26.39 \pm 0.09 }$ \\
5 & 2 18 52.26 & --5 12 48.1 & 3.117 & 2.16 & $\mathrm{25.41 \pm 0.10 }$ & -- & $\mathrm{>28.86 }$ & $\mathrm{26.84 \pm 0.15 }$ \\
6 & 2 18 51.24 & --4 59 37.8 & 3.099 & 4.44 & $\mathrm{24.40 \pm 0.04 }$ & $\mathrm{24.77 \pm 0.13 }$ & $\mathrm{26.21 \pm 0.06 }$ & $\mathrm{25.40 \pm 0.03 }$ \\
7 & 2 18 48.24 & --5 09 22.8 & 3.067 & 6.53 & $\mathrm{24.27 \pm 0.03 }$ & -- & $\mathrm{25.92 \pm 0.04 }$ & $\mathrm{25.39 \pm 0.03 }$ \\
8 & 2 18 45.14 & --4 51 43.8 & 3.137 & 3.97 & -- & $\mathrm{24.56 \pm 0.12 }$ & $\mathrm{26.84 \pm 0.13 }$ & $\mathrm{26.18 \pm 0.10 }$ \\
9 & 2 18 38.57 & --4 57 38.7 & 3.054 & 9.62 & $\mathrm{24.87 \pm 0.06 }$ & -- & $\mathrm{26.44 \pm 0.10 }$ & $\mathrm{26.21 \pm 0.10 }$ \\
10 & 2 18 36.23 & --5 08 49.8 & 3.150 & 6.45 & -- & $\mathrm{24.06 \pm 0.06 }$ & $\mathrm{25.68 \pm 0.03 }$ & $\mathrm{25.01 \pm 0.02 }$ \\
11 & 2 18 34.31 & --4 57 13.2 & 3.149 & 4.49 & -- & $\mathrm{24.32 \pm 0.08 }$ & $\mathrm{>28.86 }$ & $\mathrm{25.99 \pm 0.10 }$ \\
12 & 2 18 29.62 & --4 57 27.7 & 3.132 & 4.25 & -- & $\mathrm{24.46 \pm 0.09 }$ & $\mathrm{27.70 \pm 0.36 }$ & $\mathrm{26.80 \pm 0.21 }$ \\
13 & 2 18 24.41 & --5 05 21.5 & 3.113 & 17.29 & $\mathrm{23.90 \pm 0.02 }$ & $\mathrm{23.27 \pm 0.03 }$ & $\mathrm{26.27 \pm 0.06 }$ & $\mathrm{25.33 \pm 0.03 }$ \\
14 & 2 18 18.85 & --4 50 01.9 & 3.146 & 5.05 & -- & $\mathrm{24.32 \pm 0.09 }$ & $\mathrm{26.13 \pm 0.06 }$ & $\mathrm{25.29 \pm 0.04 }$ \\
15 & 2 18 10.41 & --4 47 18.3 & 3.127 & 3.98 & -- & $\mathrm{24.57 \pm 0.12 }$ & $\mathrm{26.70 \pm 0.10 }$ & $\mathrm{26.06 \pm 0.07 }$ \\
16 & 2 18 08.14 & --4 46 57.1 & 3.134 & 4.16 & -- & $\mathrm{24.53 \pm 0.11 }$ & $\mathrm{27.48 \pm 0.20 }$ & $\mathrm{26.28 \pm 0.08 }$ \\
17 & 2 18 07.82 & --5 11 55.5 & 3.076 & 4.39 & $\mathrm{24.47 \pm 0.04 }$ & -- & $\mathrm{26.08 \pm 0.05 }$ & $\mathrm{25.49 \pm 0.03 }$ \\
18 & 2 17 59.15 & --4 51 13.9 & 3.136 & 3.01 & -- & $\mathrm{24.72 \pm 0.10 }$ & $\mathrm{27.18 \pm 0.13 }$ & $\mathrm{26.60 \pm 0.10 }$ \\
19 & 2 17 55.07 & --5 04 12.0 & 3.062 & 3.47 & $\mathrm{25.15 \pm 0.06 }$ & -- & $\mathrm{26.42 \pm 0.06 }$ & $\mathrm{26.12 \pm 0.06 }$ \\
20 & 2 17 41.22 & --5 04 39.8 & 3.124 & 4.78 & -- & $\mathrm{24.38 \pm 0.08 }$ & $\mathrm{27.91 \pm 0.26 }$ & $\mathrm{26.52 \pm 0.09 }$ \\
\enddata
\tablenotetext{a}{Redshift measured from the Ly$\alpha$ emission line. Its error is typically less than 0.001.}
\tablecomments{The whole sample is provided in the electronic version. Upper limits are given if one object is fainter than $2\sigma$ detection.}
\end{deluxetable*}

\section{The LAE Sample at $z\approx3.1$ } \label{sec:lya_UV}

In this section, we calculate the Ly$\alpha$ redshifts of our LAEs from the spectra. We then use the narrowband and broadband photometry, together with the  redshifts, to measure spectral properties, such as UV continuum flux and slope, Ly$\alpha$ line flux and EW.

\subsection{Redshifts}

\begin{figure}[t]
\plotone{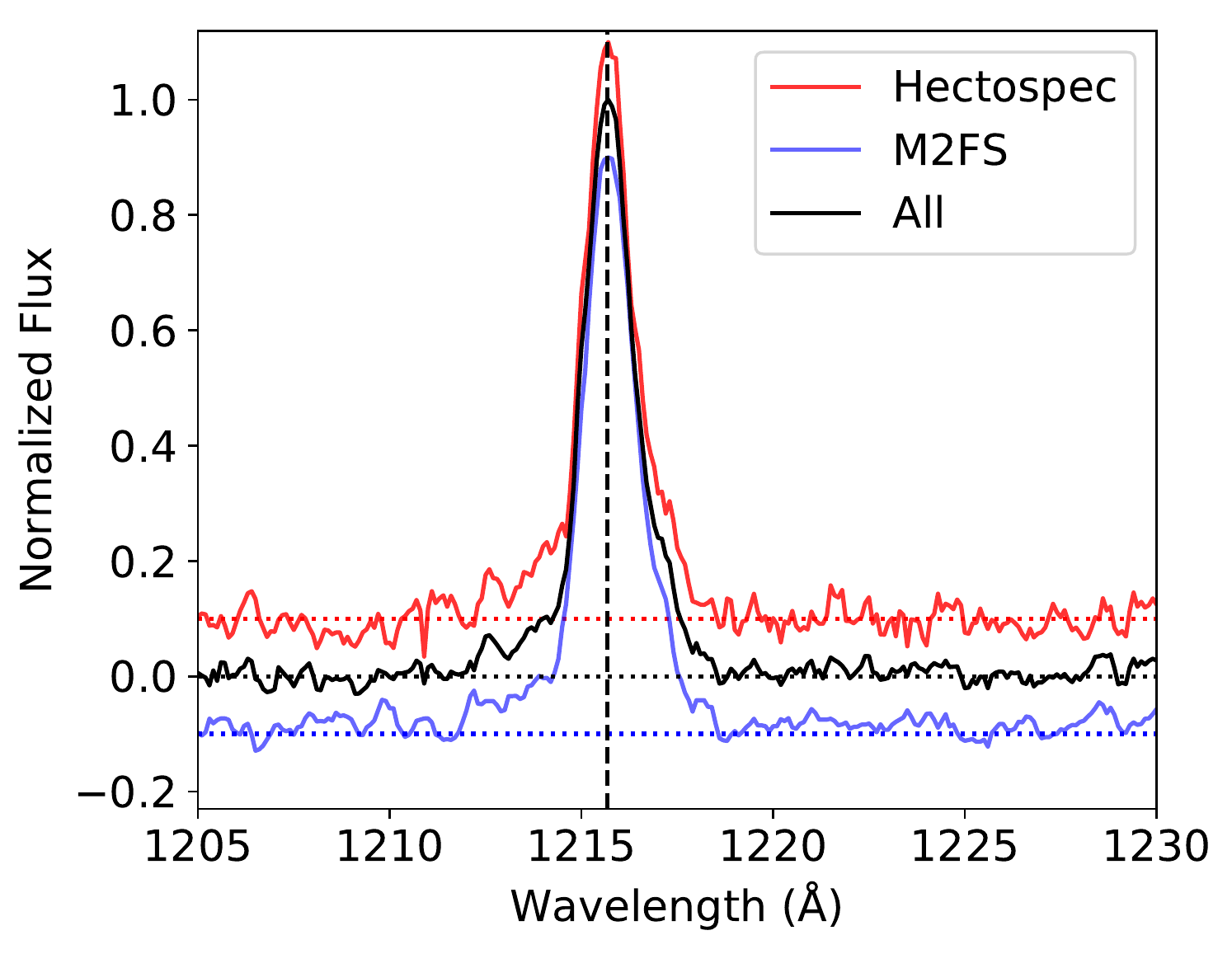}
\caption{The co-added Ly$\alpha$ line profiles. The black line represent the co-added profile of all LAEs. The blue and yellow lines represent the co-added profiles from the LAEs observed by Hectospec and M2FS, respectively. They are vertically shifted for clarity. The three profiles are consistent. \label{fig:profile}}
\end{figure}

We measure Ly$\alpha$ redshifts by fitting a composite, high-quality Ly$\alpha$ line profile to individual Ly$\alpha$ lines. We first estimate redshifts for individual Ly$\alpha$ lines based on the wavelengths of their peak flux. We then stack the spectra of all LAEs based on the individual redshifts and obtain a median Ly$\alpha$ line profile. Next, we calculate a new redshift for each LAE by fitting the median profile to its Ly$\alpha$ line. The fit was done using three parameters, the Ly$\alpha$ line peak wavelength, peak flux, and a scale factor that determines the line width. We iterate this procedure several times. The redshifts in this procedure are calculated by $z\mathrm{=\lambda_{Ly\alpha}/1215.67-1}$. The redshift distribution of all LAEs is shown in Figure~\ref{fig:redshift}. It should be noted that there is an offset between a Ly$\alpha$ redshift and its systemic redshift. The Ly$\alpha$ emission line is usually redshifted by a few hundred $\mathrm{km\: s^{-1}}$ \citep[e.g.,][]{verhamme18}.

The final median Ly$\alpha$ profile is shown in Figure~\ref{fig:profile}.
The blue and red lines represent the co-added profiles observed by Hectospec and M2FS, respectively. The black line is the final co-added profile $\mathrm{S_{Ly\alpha}}$. The three profiles are consistent with each other. They appear asymmetric, with the left side steeper than the right side, due to the ISM absorption. The bottom left part of the profile at 1212$\sim$1214 \AA\ seems to have a small bump.
\citet{hashimoto15} presented the spectra of a sample of LAEs at $z \approx 2.2$ and found that many LAEs have a similar bump blueward of systemic redshifts. Such a bump can be explained as the absorbed blue wing of the Ly$\alpha$ line \citep[e.g.,][]{barnes11}. \citet{hayes11} recently demonstrated the redshift evolution of \lya\ line profiles. Their results show that the blueshifted emission is rapidly suppressed by stochastic IGM absorption with increasing redshift, and the residual of the blue line wing looks like the small blue bump that we see in our stacked profile. This small bump can also be due to galaxy outflow. The models of \citet{chung16} showed that the blue bumps can be explained by an additional static shell of hydrogen that is associated with outflows confined to the ISM.

The composite spectrum will help us explore weaker spectral features which are otherwise too faint to see in individual spectra. We have detected the N V $\lambda$1239 line and the C IV doublet at 1550 \AA\ in the composite spectra. We will discuss this later. Note that we have used Ly$\alpha$ redshifts to stack individual spectra, which may have weakened the flux of other emission lines due to the offsets between the Ly$\alpha$ redshifts and systemic redshifts.

\subsection{Ly$\alpha$ Line and UV Continuum Flux}
\label{subsec:lya_UV_fit}

The UV continuum emission of a typical LAE in our sample is very weak, so we cannot directly measure it from the spectrum. We measure the Ly$\alpha$ line flux and UV continuum properties using the method given by \citet{jiang13}. We first build a model spectrum that includes a Ly$\alpha$ line profile and a power-law UV continuum,
\begin{equation}
f_{gal}\mathrm{=A\times S_{Ly\alpha}+B\times\lambda ^{\beta}},
\label{eq:f_gal}
\end{equation}
where A is the Ly$\alpha$ peak flux density, $\mathrm{S_{Ly\alpha}}$ is the co-added Ly$\alpha$ line profile shown in Figure~\ref{fig:profile}, B is a scale factor of the UV continuum, and $\beta$ is the UV continuum slope. A and B are in units of $\mathrm{erg\, s^{-1}\,\AA ^{-1}\,cm^{-2}}$.

We know that the wavelength range of the spectrum blueward of Ly$\alpha$ cannot be described by a power law due to the ISM and IGM absorption. We apply an average scale factor $C$ to this part of the spectrum. The scale factor $C$ is estimated using the two narrowband photometry in NB497 and NB503. We first select a sample of $z>3.14$ LAEs that have photometric measurement in NB497. The NB497 filter barely covers Ly$\alpha$ for these LAEs, so the NB497 photometry represents the continuum flux blueward of Ly$\alpha$. We then select another sample of $z<3.09$ LAEs that have photometric measurement in NB503. The NB503 filter barely covers Ly$\alpha$ for these LAEs, so the NB503 photometry represents the continuum flux redward of Ly$\alpha$. We scale all LAEs in the two samples to the same continuum level, and calculate the median flux ratio of NB497 to NB503. This ratio is the scale factor $C$, which is roughly 0.61. Based on the model given by \citet{madau95}, we get a similar absorption at the blue side of \lya\ of $C\sim0.62$.

\begin{figure}[t]
\plotone{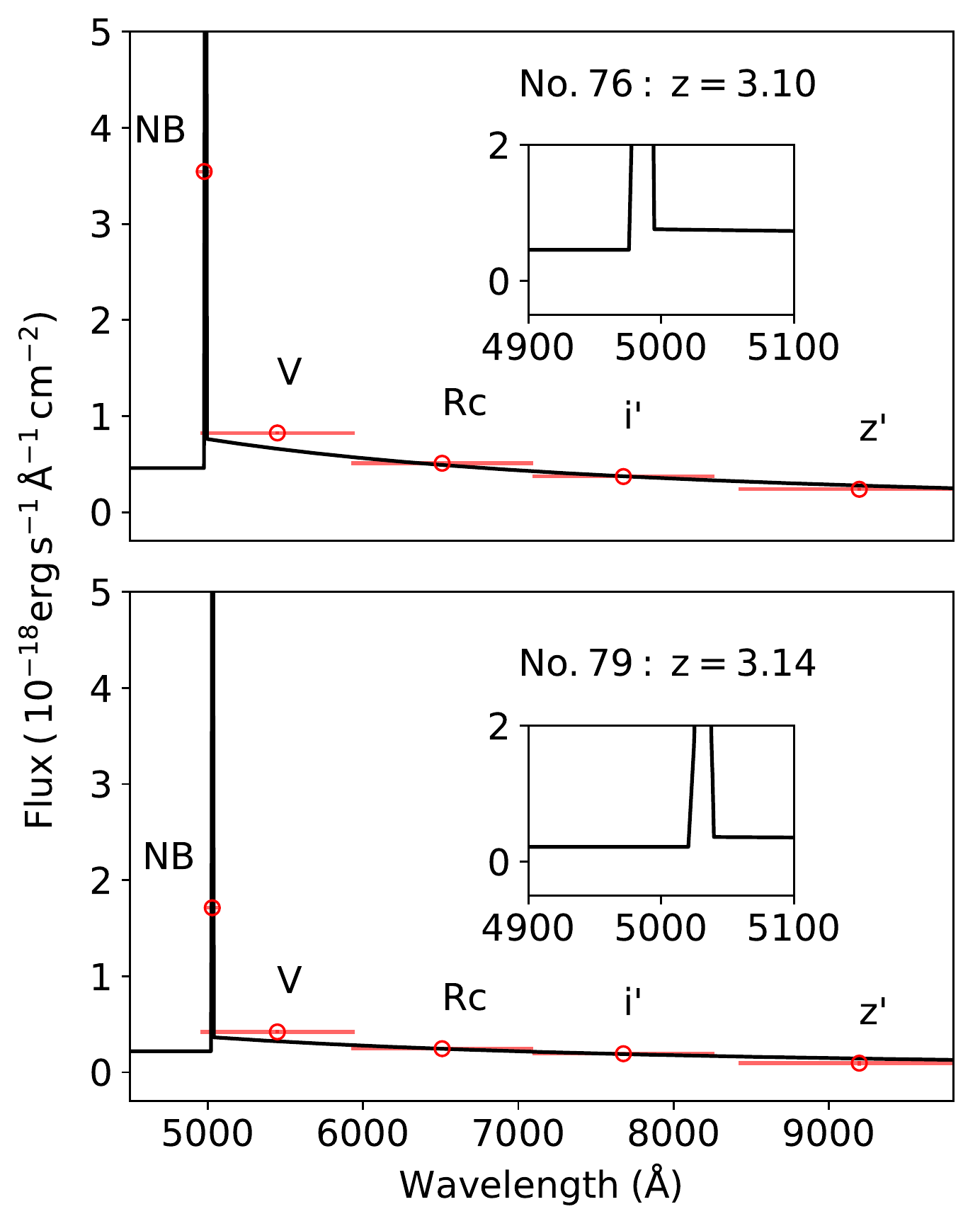}
\caption{Demonstration of our measurement of the Ly$\alpha$ and continuum properties. We show two examples in the upper and lower panels.
The red circles show the photometric data points with the horizontal bars indicating the wavelength coverage of each band. The photometric errors are very small and invisible in the figure. The solid profile in each panel is the best-fitted model spectrum that consists of a Ly$\alpha$ line and a power-law continuum. The insets show the regions around Ly$\alpha$. \label{fig:flux}}
\end{figure}

\begin{figure*}[t]
\plotone{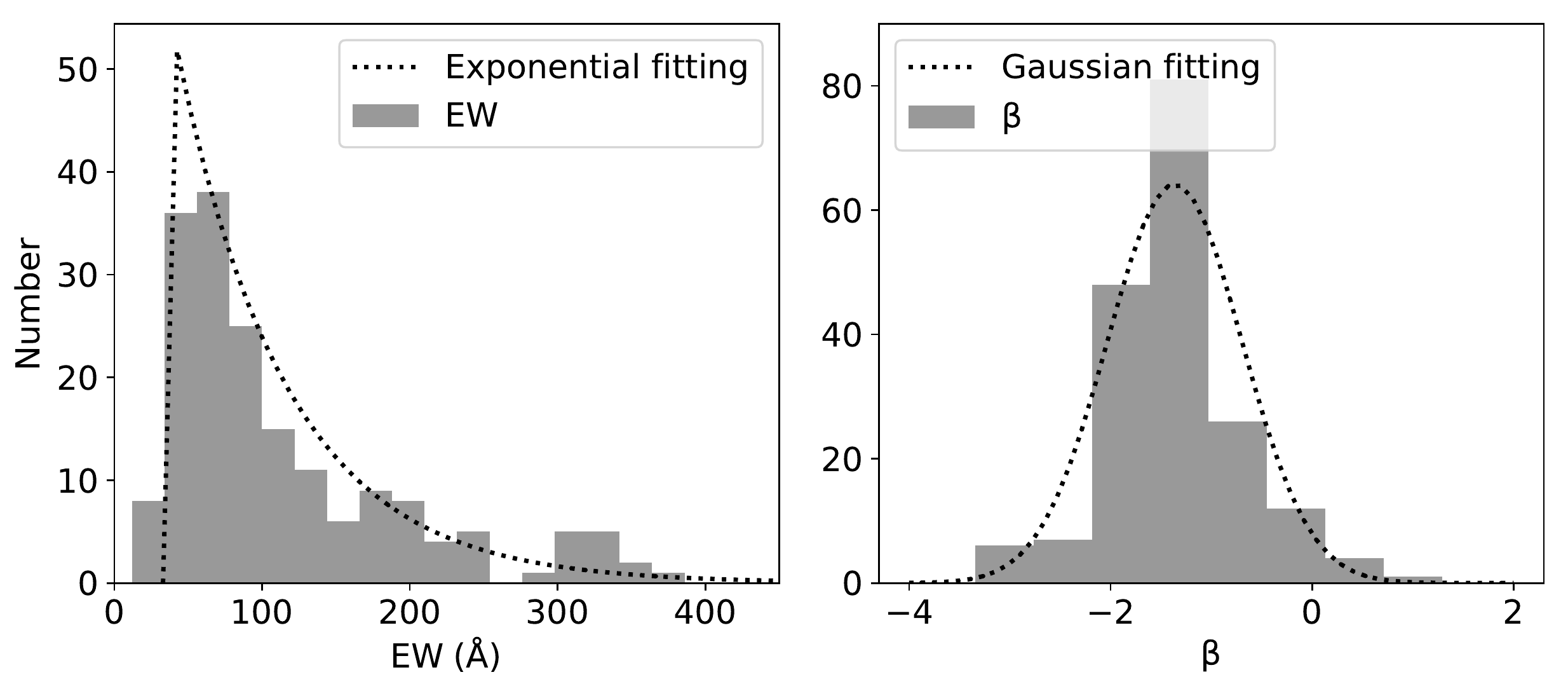}
\caption{Ly$\alpha$ EW distribution (left panel) and UV slope $\beta$ distribution (right panel) of the LAEs in our sample. In the left panel, the black dotted line represent a best-fitted exponential model. In the right panel, the black dotted line represents a best-fitted Gaussian model. \label{fig:ew0_beta}}
\end{figure*}

In our calculation, the UV continuum properties (B and $\beta$ in Equation~\ref{eq:f_gal}) are mainly constrained by the photometry in the $V$, $Rc$, $i'$, $z'$ bands. The Ly$\alpha$ flux is mainly constrained by the narrowband photometry. We build a grid of A, B and $\beta$ values and produce a large amount of model spectra following Equation~\ref{eq:f_gal}. We then convolve these spectra with the filter response curves to compute photometry in each band. Finally, by comparing the calculated values with the observed values, we obtain the $\chi^2$-optimized A, B, and $\beta$ values.
Note that our measurement is not sensitive to $C$, because the narrowband photometry is dominated by Ly$\alpha$, and the two narrow bands are at the edge of the $V$ band (we did not use the $B$ band). With different $C$, the Ly$\alpha$ flux changes by $<5\%$. We actually iterate the above procedure several times, because the measurement of $C$ requires the continuum level redward of Ly$\alpha$. We show two examples in Figure 7.

From the redshifts and the best-fit B, $\beta$, and A values, we measure LAE spectral properties including the Ly$\alpha$ line flux and luminosity, UV continuum luminosity $\mathrm{L_{1500}}$, and Ly$\alpha$ EW. The left panel of Figure~\ref{fig:ew0_beta} shows the histogram of the EWs in our sample. The EW distribution of LAEs has been found to have an exponential form $\mathrm{dN/dEW\propto exp(-EW/W_0)}$, where $\mathrm{W_0}$ is a scale length.
We fit an exponential function to the observed distribution using an MCMC approach given by \citet{santos20}. We set a lower limit of 45 \AA\ for EW and an upper limit of 240 \AA\ \citep{charlot93}. The scale length that we obtain from our sample is $74.3 \pm 9.2$ \AA.
{\color{blue}
There are many studies of $\mathrm{W_0}$ in the literature \cite[e.g.,][]{gronwall07,guaita10,kashikawa11,ciardullo12,wold14,zheng14,wold17,hashimoto17,jung18,shibuya18a}.} In these studies, $\mathrm{W_0}$ is roughly within a range of $60\sim100$ \AA\ at $0.3<z<6$.
In addition, $\mathrm{W_0}$ tends to be larger at higher redshift towards $z\sim6$ because of lower metallicity and/or less dust. At $z>6$, \lya\ EW values become smaller due to the IGM absorption.
Our result of  $\mathrm{W_0}=74.3$ \AA\ is consistent with previous measurements at similar redshifts. For example, \citet{gronwall07} measured a scale length of $\sim$76 \AA\ and \citet{ciardullo12} obtained a scale length of $\sim$70 \AA\ at $z\sim3.1$.

The rest-frame UV-continuum slope $\beta$ provides important information to constrain stellar populations in galaxies.
In the right panel of Figure~\ref{fig:ew0_beta}, we plot the distribution of $\beta$ in our sample. The average slope, measured from a Gaussian fit, is $\beta=-1.38\pm0.15$ with a standard deviation of $0.69\pm0.06$. The median slope is  $\beta = -1.43$.
Our $\beta$ distribution is reliable, not only because these galaxies are spectroscopically confirmed, but also because our target selection criteria did not reply on UV-continuum slopes.

\section{Ly$\alpha$ Luminosity function} \label{sec:lf}

\subsection{Completeness of the sample}
\label{subsec:lf-completeness}

The Ly$\alpha$ LF is a fundamental statistical property of LAEs. In order to measure LF, we need to correct the incompleteness of the sample.
Sample incompleteness usually originates from four aspects, source detection in imaging data, galaxy candidate selection, spectroscopic observations, and  LAE identification. In this section, we will provide the details about the correction of our sample incompleteness.

\subsubsection{Source Detection}
\label{sec:detection_ratio}

\begin{figure}[t]
\includegraphics[width=8.5cm]{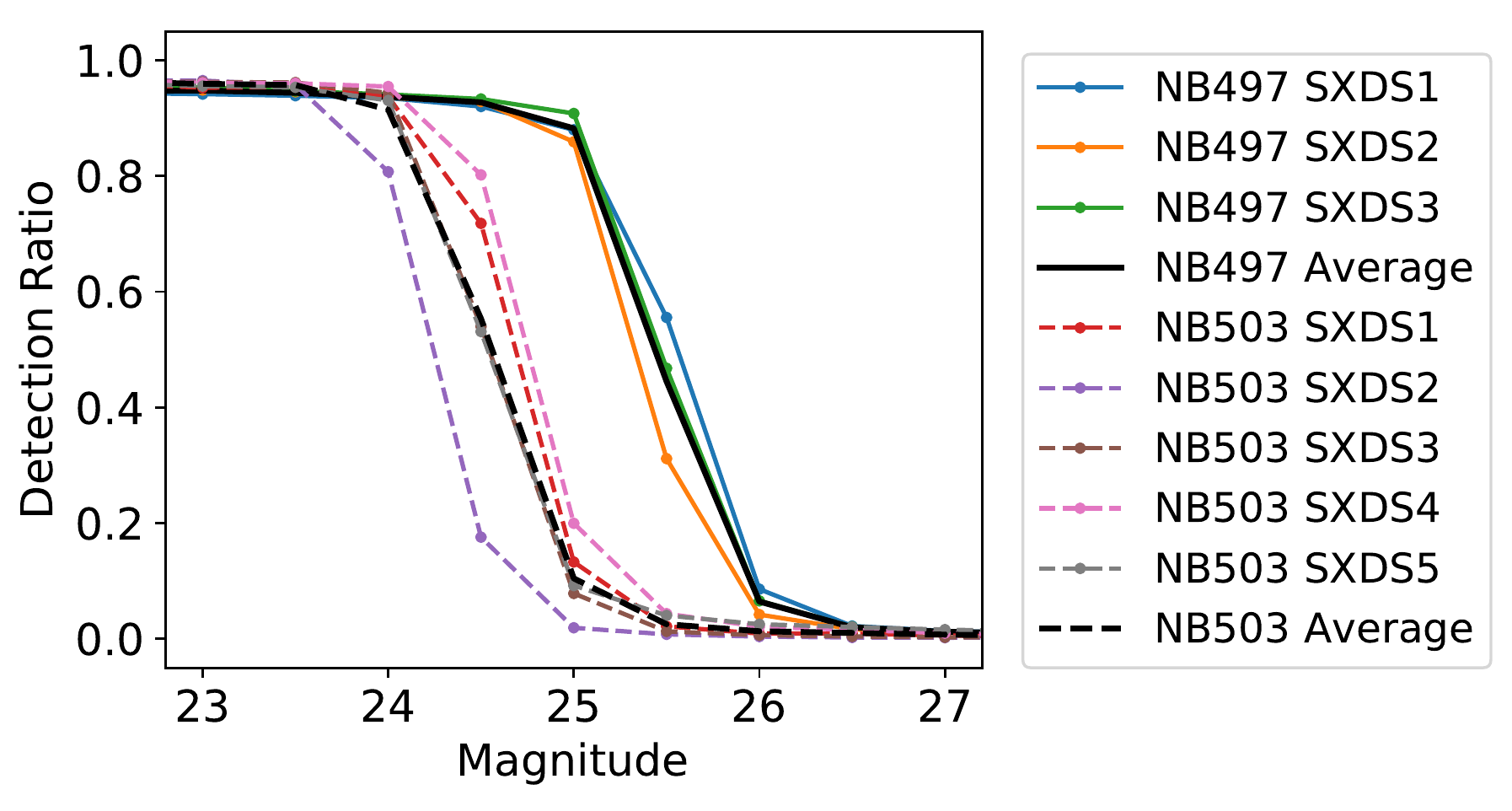}
\caption{Source detection rates of mock LAEs in NB497 and NB503. The average detection ratios in NB497 and NB503 are illustrated by the black solid line and black dashed line, respectively. \label{fig:detection_ratio}}
\end{figure}

The first incompleteness came from the source detection in our imaging data. We estimate this incompleteness using a Monte Carlo simulation, i.e.,
we calculate the recovery percentage of randomly distributed, artificial sources in our images.
We first produce a median image of LAEs by co-adding narrowband images of all LAEs in our sample. This median image represents the typical morphology of our LAEs. We then randomly put 100 these mock LAEs in the NB497 and NB503 images, and run SExtractor to detect these objects. We also require that these objects should be in clean regions in the $B$- and $V$-band images. We repeat this procedure 1000 times and calculate recovery rates.
The result is shown in Figure~\ref{fig:detection_ratio}. On average, the NB497-band images are about 0.9 mag deeper than the NB503-band images.

\subsubsection{Color Selection}
\label{subsubsec:imcompleteness-color}

\begin{figure*}[ht!]
\plotone{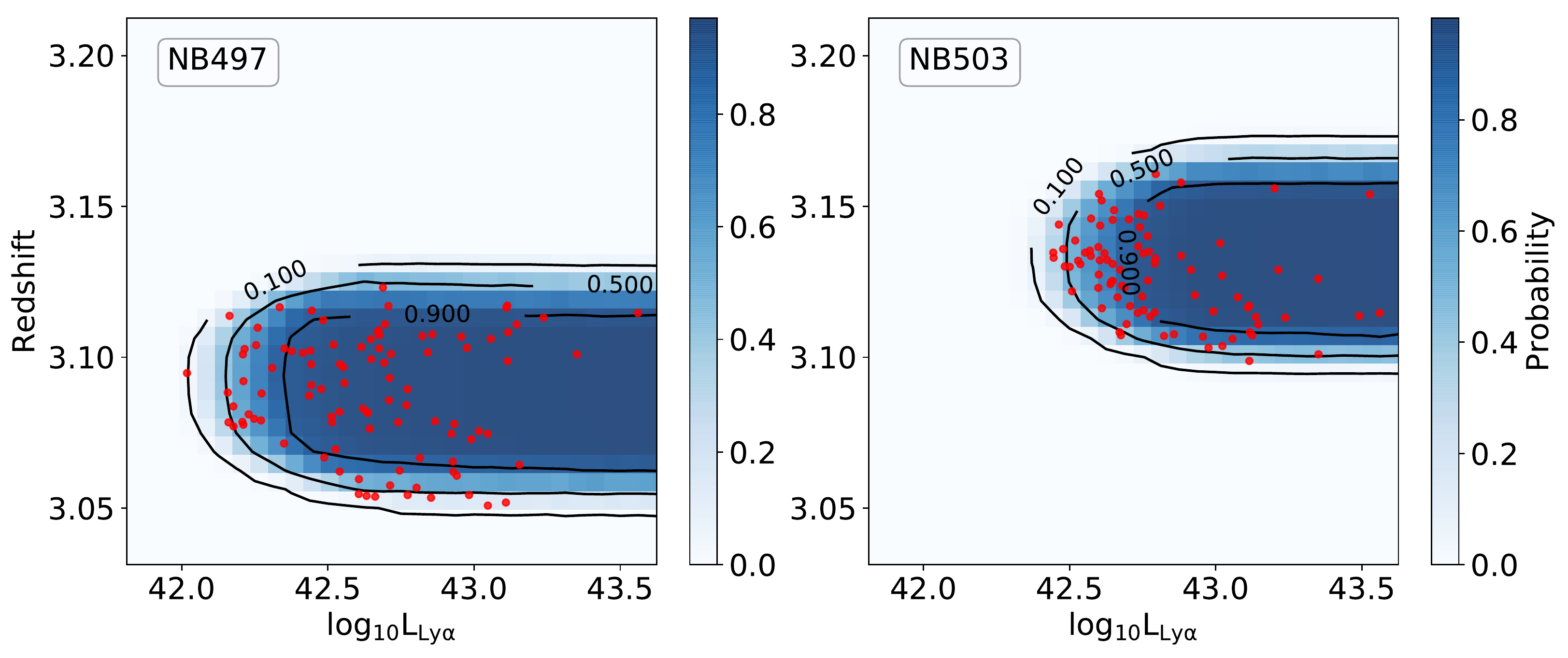}
\caption{Selection probability as a function of Ly$\alpha$ luminosity and redshift. The contours represent the probabilities of 0.1, 0.5, and 0.9. The probability is shown by the color bar. The red dots represent the LAEs in our sample. \label{fig:completeness}}
\end{figure*}

The second incompleteness came from the color selection, i.e., the probability that a LAE meets our color selection criteria. We run a simulation to estimate this incompleteness. We first generate simulated LAE spectra following Equation~\ref{eq:f_gal}. The UV slope $\beta$ of the simulated spectra has the same Gaussian distribution as shown in the right panel of Figure~\ref{fig:ew0_beta}, and the Ly$\alpha$ EW has the same exponential distribution as shown in the left panel of Figure~\ref{fig:ew0_beta}. We further assume that the distributions of EW and $\beta$ do not change with Ly$\alpha$ luminosity. The assumed distributions have very small impact on our results, because our target selection does not rely on UV continuum slope and the Ly$\alpha$ flux dominates the narrowband photometry.

We construct a grid of $\mathrm{log_{10}L_{Ly\alpha}}$ and $z$ in the ranges of $3.00<z<3.25$ and $\mathrm{42.00<Log_{10}L_{Ly\alpha}<44.00}$. The step sizes are $\mathrm{\Delta Log_{10}L_{Ly\alpha}=0.01}$ and $\Delta z=0.01$. For each pair of [$\mathrm{log_{10}L_{Ly\alpha}}$, $z$], we generate 10,000 simulated LAE spectra that follow the above EW and $\beta$ distributions. For each spectrum, we calculate its $B$, $V$, $BV$,and narrowband photometry. We also add photometric errors that follow the magnitude-error relations from real images. We then feed this spectrum to our selection criteria to check if this LAE can be selected. The selection completeness for this [$\mathrm{log_{10}L_{Ly\alpha}}$, $z$] pair is the probability that the 10,000 simulated LAEs are selected.

The final results is shown in Figure~\ref{fig:completeness}. In this figure, we also include the detection completeness shown in Figure~\ref{fig:detection_ratio}. The figure shows high completeness for LAEs in the both narrow bands. The mean completeness is $\sim 81\%$ for all LAEs. The $50\%$ completeness limit reaches $\mathrm{log_{10}L_{Ly\alpha} \sim 42.2}$ for NB497, and $\sim$ 42.5 for NB503.

\subsubsection{Spectroscopic Observations}
\label{subsubsec:imcompleteness-specobs}

The third incompleteness came from our spectroscopic observations, namely, the fraction of LAE candidates that have been spectroscopically observed.
In our program, we got 345 LAE candidates in our survey area, and 265 of them were observed by Hectospec or M2FS. Among the candidates that were not spectroscopically observed, a small fraction (16) of them had been identified by \citet{ouchi08}. The others were not observed due to fiber collision. Overall $\sim$81\% of the LAE candidates were observed spectroscopically: the fraction observed by M2FS is 96\% and the fraction observed by Hectospec is 71\%. Therefore, two correction factors 1/0.96 and 1/0.71 are applied, respectively.

\subsubsection{LAE Identification}
\label{subsubsec:imcompleteness-identification}

The fourth incompleteness came from the LAE identification. Our M2FS spectral data reach a depth of $\sim$25.8 mag in the narrow band, and the Hectospec data reach a depth of $\sim$25.4 mag. They are deep enough to identify Ly$\alpha$ emission lines down to our sample limits. In addition, there are no obvious OH skylines at $\sim$5000 \AA. Therefore, we assume that this completeness is nearly 100\%.

\begin{figure}
\plotone{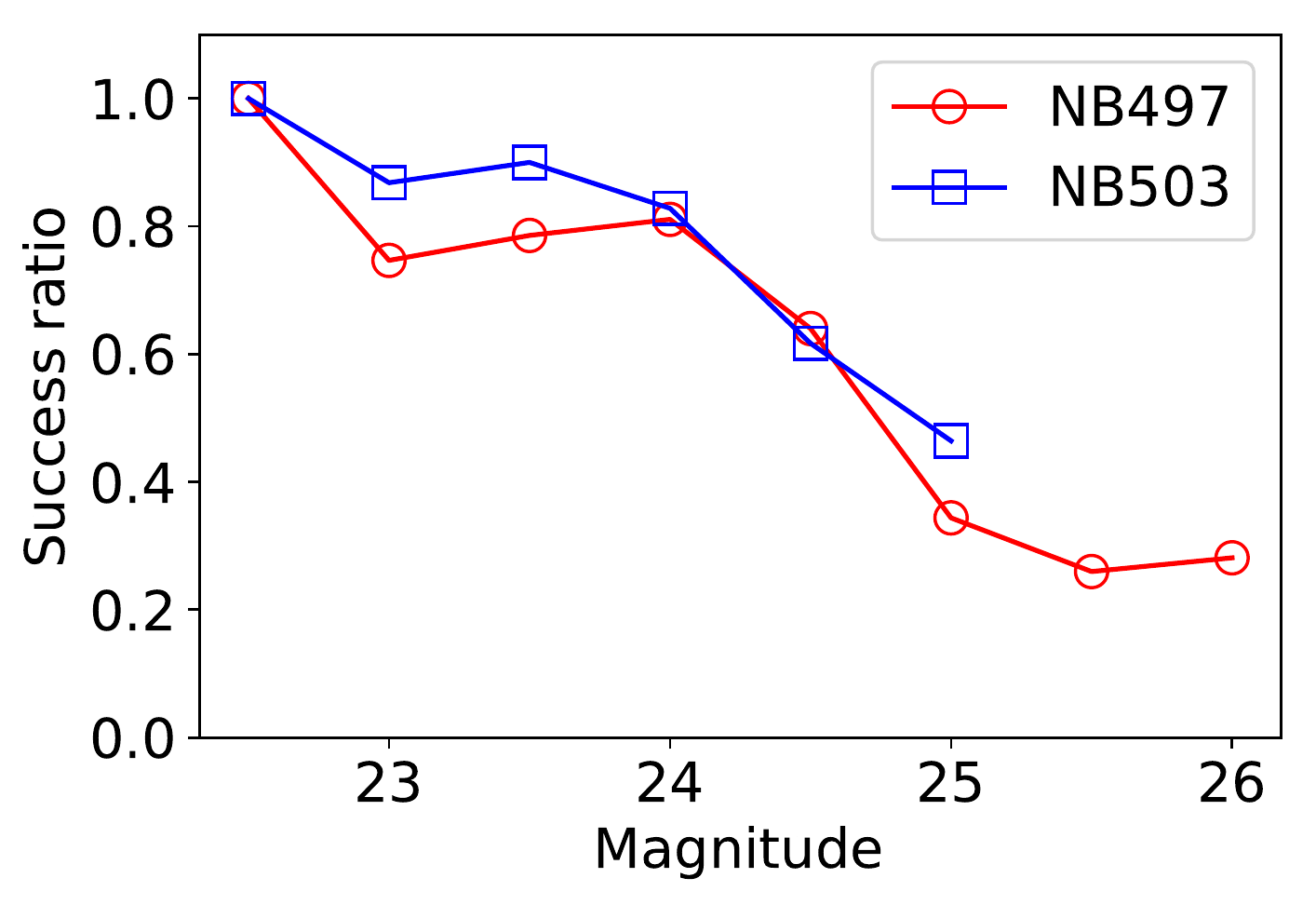}
\caption{ The spectroscopic success fractions of LAEs as a function of narrow band magnitude. The success fractions of NB497 and NB503 are shown by red and blue line, respectively. \label{fig:success_ratio}}
\end{figure}

Figure~\ref{fig:success_ratio} shows the spectroscopic success rates as a function of magnitude in the two narrow bands, e.g., the fractions of the confirmed LAEs in the two candidate samples. The fractions reach 100\% for the most luminous targets, and decline towards fainter magnitudes. This also means that the contamination rates in the candidates increase towards fainter magnitudes. The majority of the contaminants do not show a detectable emission line in the expected wavelength rage ($\sim$5000 \AA). As we mentioned above, our spectroscopic observations are deep enough (by design) to identify a line feature at $\sim$5000 \AA\ down to our sample limit, a target without a line detections was reliably classified as a non-LAE. The success rates at NB$>25$ are significantly lower, because we included less promising, faint candidates (Section.~\ref{subsec:obs-selection}).

\subsection{The $\mathrm{1/V_{a}}$ Estimate}
\label{subsec:lf-vmax}

\begin{figure*}
\plotone{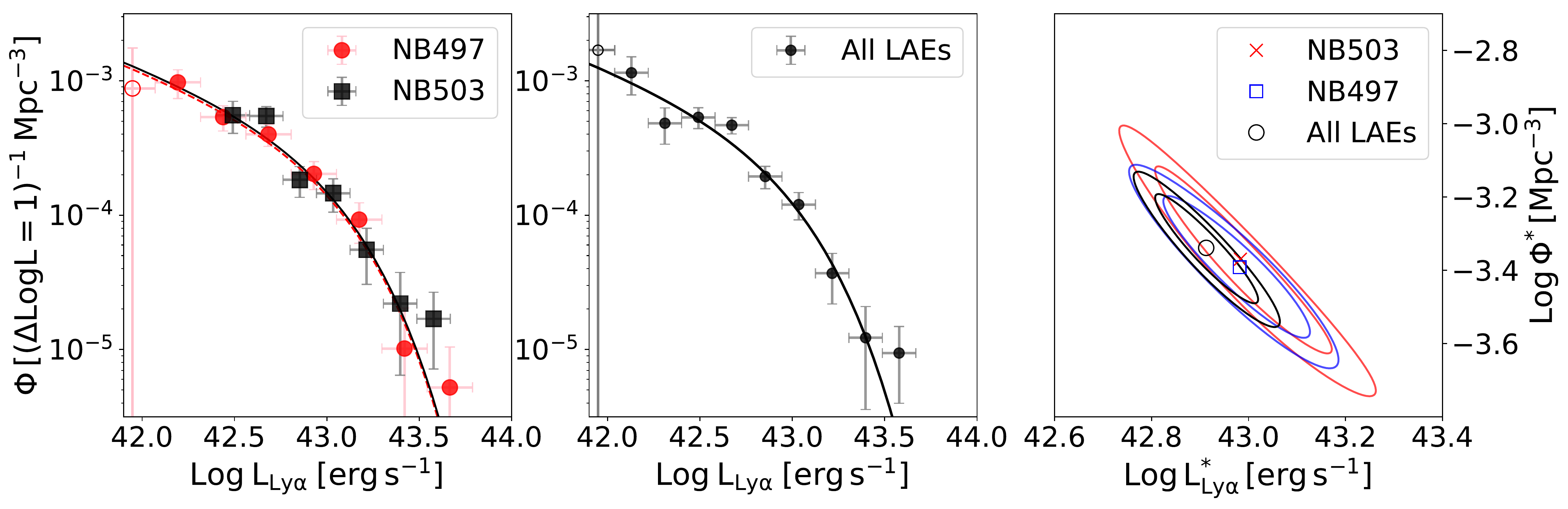}
\caption{ Ly$\alpha$ LFs. Left: Ly$\alpha$ LFs from the two LAE samples in NB497 and NB503, respectively. The black and red symbols with error bars represent the binned LFs and the curves represent the best model fits to the Schechter function. Middle: The binned LF and the best model fit for the whole sample. Right: Contours of fitting parameters $L^*$ and $\Phi^*$ at a fixed $\alpha=-1.6$. We show the 68\% and 90\% confidence regions.  \label{fig:LF-NB}}
\end{figure*}

We use the $\mathrm{1/V_{a}}$ method \citep[e.g.,][]{avni80} to estimate a binned Ly$\alpha$ LF for our LAE sample. The LAEs are grouped into different luminosity bins. Redshift evolution is ignored. The cosmic volume available to discover a LAE with Ly$\alpha$ luminosity $L'$ and redshift $z'$ is
\begin{equation}
V_{a}=\frac{1}{\Delta \mathrm{log}(L)}\int_{\Delta \mathrm{log}(L)}^{ }\int_{\Delta z}^{ }p(L',z')\, \frac{dV}{dz}\, dz\, d(\mathrm{log}L),
\end{equation}
where $p(L,z)$ is a probability function of $L$ and $z$ that combines all incompleteness mentioned in Section.~\ref{subsec:lf-completeness}, and $\Delta z$ is the redshift range determined by the narrowband filters.

The differential Ly$\alpha$ LF $\Phi (L)$ is the spatial density of galaxies per luminosity bin $\Delta \mathrm{log}L$. In a given bin $\Delta \mathrm{log}L$ centered at $L_i$, $\Phi (L_i)$ is given by
\begin{equation}
\Phi(L_i)=\frac{1}{\Delta \mathrm{log}L} \sum_{j}\frac{1}{V_{a,j}},
\end{equation}
where $i$ denotes the luminosity bin number and $j$ denotes the galaxy number. The uncertainty is written as
\begin{equation}
\sigma[\Phi(L_i)]=\frac{1}{\Delta \mathrm{log}L} [\sum_{j}(\frac{1}{V_{a,j}})^2]^{1/2}.
\end{equation}

Left panel of Figure~\ref{fig:LF-NB} shows our results for NB497 and NB503 separately. The LF of the NB497 sample reaches $\L_{Ly\alpha} \sim \mathrm{10^{42}\,erg\,s^{-1}}$. Its faintest bin consists of one LAE with a very low completeness ($<0.1$) as seen in Figure~\ref{fig:completeness}, so we exclude this bin in the following analyses. The NB503 sample is slightly shallower. The LFs from the two samples are consistent. We present the Ly$\alpha$ LF for the whole sample in the middle panel of Figure~\ref{fig:LF-NB}. In Figure~\ref{fig:LF-all}, we also compare our results with those from the literature.

In order to parameterize the LFs, we fit the binned Ly$\alpha$ LFs using a Schechter function
\begin{equation}
\Phi (L)d\mathrm{log}L=ln10\Phi ^*(\frac{L}{L^*})^{\alpha+1}e^{-L/L^*}d\mathrm{log}L,
\end{equation}
where $\Phi ^*$, $L^*$ and $\alpha$ are the characteristic number density, characteristic luminosity, and faint end slope, respectively \citep{schechter76,drake17}. We fit the Schechter function with $\chi^2$ statistics \citep[e.g.][]{malhotra04,zheng16}.
Our data are not deep enough to constrain the faint end slope $\alpha$, so we try a series of $\alpha$ values from --1.0 to --2.0 with a step of 0.1.
For each $\alpha$ value, we perform a $\chi^2$ fit. We find the minimal $\chi^2$ at $\alpha=-1.6$ (this $\alpha$ is consistent with many previous studies). When $\alpha$ is fixed at $-1.6$, the best-fit values of the other two model parameters are $\Phi ^* = 10^{-3.34^{+0.06}_{-0.09}} \; \rm Mpc^{-3}$ and $L^{*}=10^{42.91^{+0.13}_{-0.11}} \; \rm erg\;s^{-1}$. The fitting result for the whole sample is shown in the middle panel of Figure~\ref{fig:LF-NB}. We also perform LF measurements for the two narrow bands with the same $\alpha$ value. The results are shown in th left panel of Figure~\ref{fig:LF-NB}. We examine the likelihood contours in the $L^*$ - $\Phi ^*$ space in the right panel of Figure~\ref{fig:LF-NB}. The 68\% and 90\% confidence levels for $\Phi ^*$ and $L^*$ are plotted.

At the brightest end of the LF, the observed data point is above the best-fit Schechter function (by $>1\sigma$). Such a lift or bump in the bright-end LF has been reported previously \cite[e.g.][]{hayes10,blanc11}. It has been claimed that this lift can be (partly) due to AGN contribution \citep[e.g.,][]{konno16,wold17}. Our sample is spectroscopically confirmed. We will argue in Section 6.3 that AGN contribution should be small in our LAEs, but we are not able to rule out a small AGN contribution. Since the detection of the density excess is tentative with a large uncertainty, we will explore more possibilities.

\section{Discussion} \label{sec:discussion}

\subsection{Comparison with Previous Studies}

\begin{figure*}
\plotone{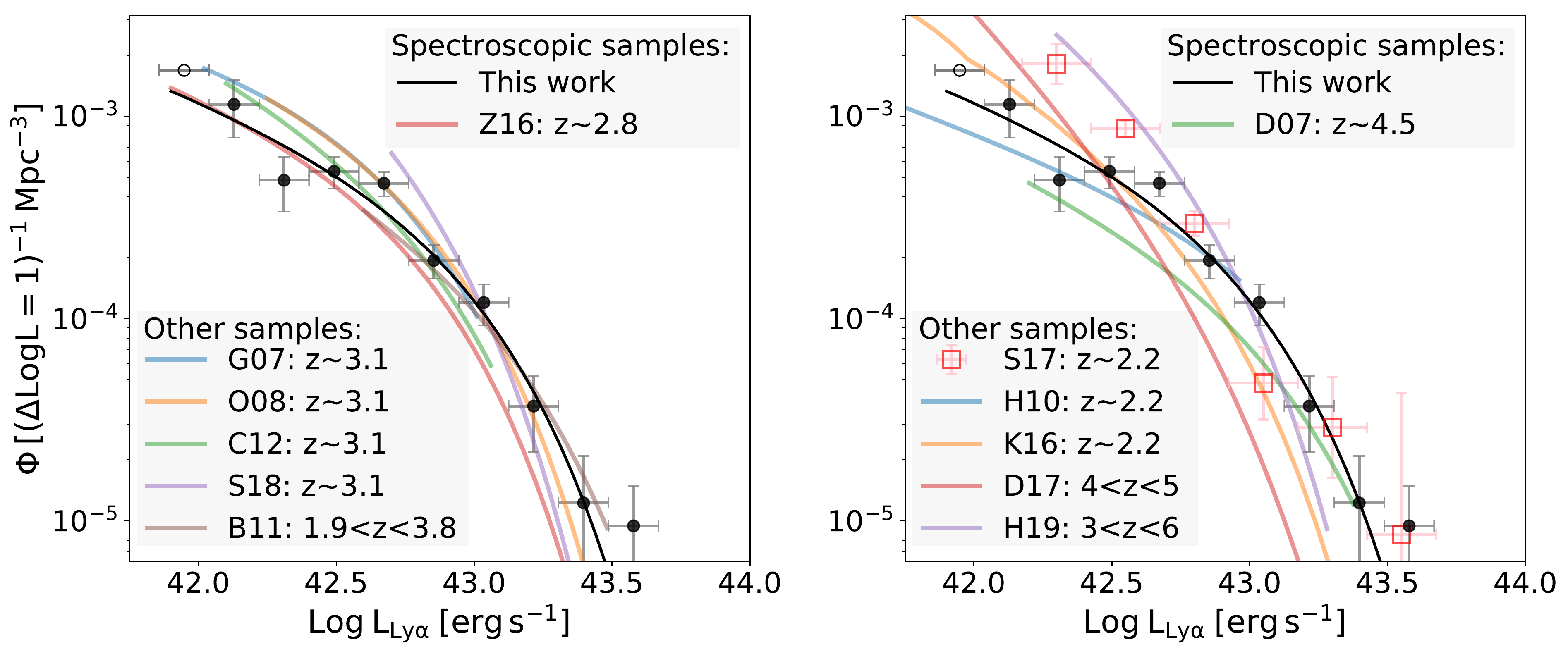}
\caption{\lya\ LF and its redshift evolution. The black dots with error bars represent the binned LF derived from our sample and the black curve represents the best model fit. Left: Comparison of our LF with the results at similar redshift $z\sim3$ in the literature  (G07: \citealt{gronwall07}; O08: \citealt{ouchi08}; C12: \citealt{ciardullo12}; Z16: \citealt{zheng16}; S18: \citealt{sobral18}; B11: \citealt{blanc11}; D17: \citealt{drake17b}). Right: Comparison with the results at different redshifts in the literature. (D07: \citealt{dawson07}; H10: \citealt{hayes10}; K16: \citealt{konno16}; D17: \citealt{drake17b}; S17: \citealt{sobral17}; H19: \citealt{herenz19}).  \label{fig:LF-all}}
\end{figure*}

Previous studies have shown little evolution of the Ly$\alpha$ LF from $z \sim 3$ to 5 \citep[e.g.,][]{ouchi08,cassata11,ciardullo12}. In this section, we compare our results with those from the literature
\citep{dawson07,gronwall07,ouchi08,bacon10,blanc11,ciardullo12,konno16,zheng16,drake17b,sobral18,herenz19}.
A few of these studies are (partly) based on spectroscopically confirmed LAE samples, including samples from blind spectroscopic surveys using IFU facilities. Other studies are based on photometrically selected LAE samples.

The comparison is shown in Figure~\ref{fig:LF-all}. In the left panel of Figure~\ref{fig:LF-all}, we compare our LF with previous results at the similar redshift $z \sim 3.1$. We can see that our LF agrees well with the previous results.
At the bright end, our LF is slightly higher than \citet{zheng16} and \citet{sobral18}, but still within the $1\sigma$ range.
In the right panel of Figure~\ref{fig:LF-all}, we compare our LF with previous results at other redshifts. It is not straightforward to explain the comparison, as different studies used different target selection criteria, observing strategy, etc. In addition, most studies were based on photometric samples. Nevertheless, our LF is generally consistent with these previous results. At the bright end, our LF is well consistent with the \citet{dawson07} and \citet{sobral17} results, but slightly higher than the other results. The \citet{dawson07} sample was spectroscopically confirmed.
\citet{konno16} found that their \lya\ LF at the bright end is significantly higher than a Schechter function. They claimed that this excess originated from the contribution of AGNs. \citet{sobral17} also found a density excess at the bright end of their \lya\ LF at $z\sim2.2$. After they removed potential contamination, their bright-end density is consistent with our result (Figure~\ref{fig:LF-all}).

At the faint end, our LF agrees well with most previous studies \citep[e.g.,][]{gronwall07,ciardullo12,konno16}, but notably lower than the LFs based on VLT MUSE \citep{drake17b,herenz19}. It is difficult to directly compare the MUSE results with other results because of the totally different target selection methods. \citet{herenz19} took the extended nature of \lya\ emission into account when constructing their selection functions. They argued that the assumption of compact point sources for LAEs would lead to a biased LF near the completeness limit. The other studies did not consider the extended emission.

\subsection{Influence of Cosmic Variance}

\begin{figure}[t!]
\plotone{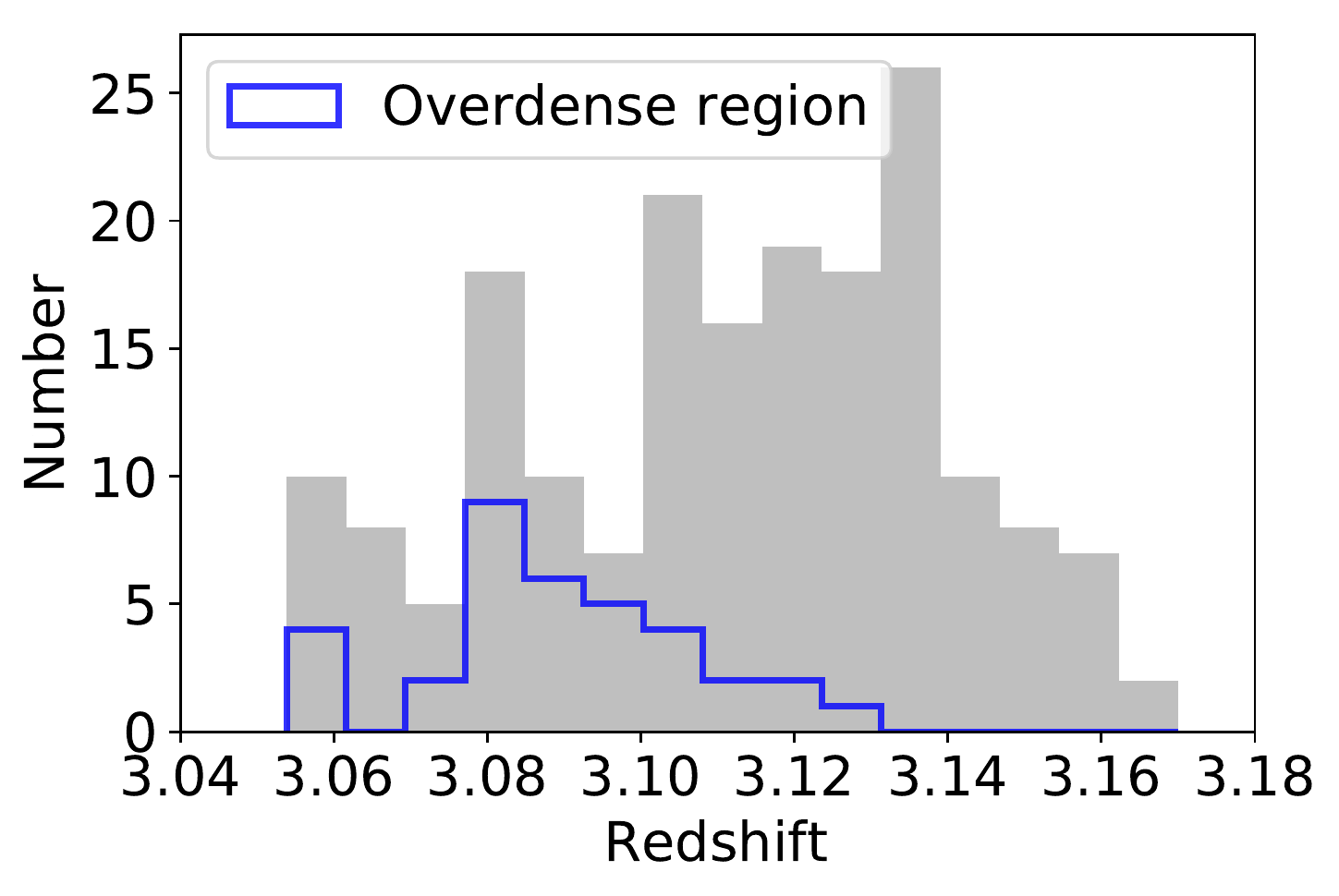}
\caption{Redshift distribution of the LAEs in the overdense region, compared with the redshift distribution of all LAEs in our sample. \label{fig:z-overdensity}}
\end{figure}

\begin{figure}[t]
\plotone{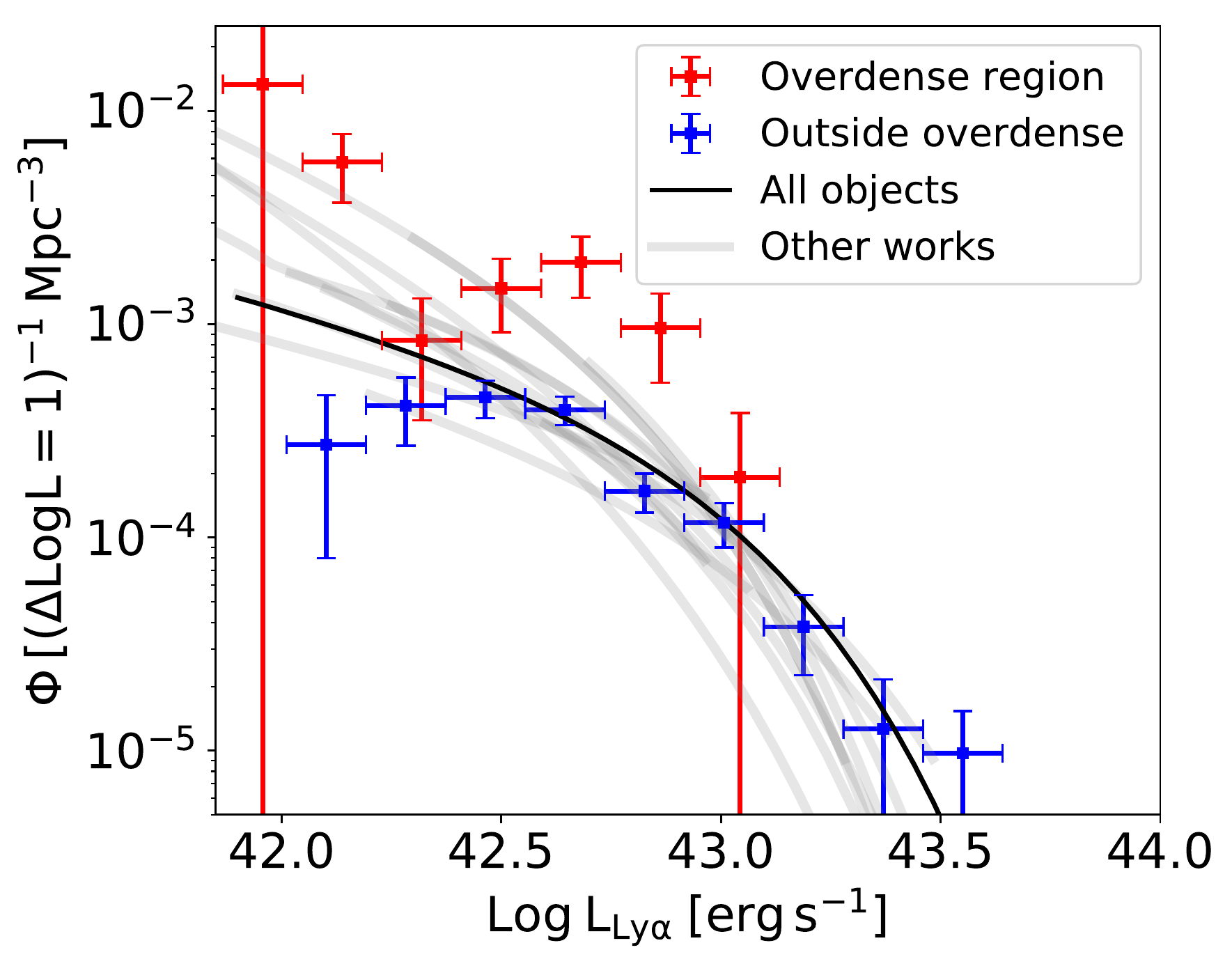}
\caption{LF of the overdense region. The red and blue dots with error bars represent the binned LFs in the overdense region in NB497 and all LAEs outside the overdense region, respectively. The LF of all LAEs is shown by the black line. The data points of the binned LFs have been slightly shifted horizontally for clarity. The results from other work (as is shown in  Figures~\ref{fig:LF-NB}, \ref{fig:LF-all}) are shown in grey. \label{fig:LF-overdensity}}
\end{figure}

LAEs are commonly used to trace large scale structures at high redshift \citep[e.g.][]{steidel00,hayashino14,cai17a,cai17b,jiang18}.
As we mentioned earlier, there is an overdense region of LAEs detected in NB497, shown in the left panel of Figure~\ref{fig:SXDS}. The projected area is about $\sim0.2\times 0.2\; \rm deg^2$. The redshift distribution of this region is plotted in Figure~\ref{fig:z-overdensity}. The median redshift is $z\approx$ 3.085. Meanwhile, we notice that the region immediately outside of the overdense region is apparently underdense. We analyze the influence of cosmic variance on our results. We calculate binned Ly$\alpha$ LFs for the overdense region and the region outside of the overdense region, and compare them with the LF of the whole LAE sample. The results are shown in Figure~\ref{fig:LF-overdensity}.

The LF (the red dots) of the overdense region is significantly higher than the LF (the black curve) of the whole sample. By comparing the two LFs, the LAE overdensity in the overdense region is $\sim4.14$. Due to the small area coverage, the overdense region lacks of very luminous LAEs, as seen in Figure~\ref{fig:LF-overdensity}. On the other hand, the LF (the blue dots) outside the overdense region is slightly lower than the LF of the whole sample in the fainter half range, mainly due to the much lower LAE density in the underdense region mentioned above. The overall effect is that the overdense and the underdense regions roughly cancel out, so that the LF from the NB497 sample agrees well with the LF from the NB503 sample (Figure~\ref{fig:LF-NB}).

The area coverage of our sample is larger than those of previous spectroscopic surveys of LAEs at $z\approx3$. It can largely reduce the influence of cosmic variance. This advantage is clearly demonstrated above.

\subsection{Detection of \nv\ and \civ} \label{subsec:discussion-nv}

Previous studies have detected UV emission lines in the spectra of individual, bright galaxies or the combined spectra of star-forming galaxies at $z\sim3$ \citep[e.g.,][]{shapley03,cassata13,zheng16,nakajima18,fevre19}. The typical lines are \civl, \heiil, \oiiiIl, and \ciiil. For example, \citet{cassata13} combined a sample of \heii\ emitters at $2<z<4.6$ and detected the \heii\ and \ciii\ emission lines in their composite spectra. \citet{zheng16} stacked a sample of LAEs at $z\approx2.8$ and detected \ciii. \citet{fevre19} combined a sample of \ciii\ emitters at $2<z<3.8$ and detected all the lines mentioned above in their different subsamples. They even detected \nvli\ that has a very high ionization potential.

\begin{figure}
\plotone{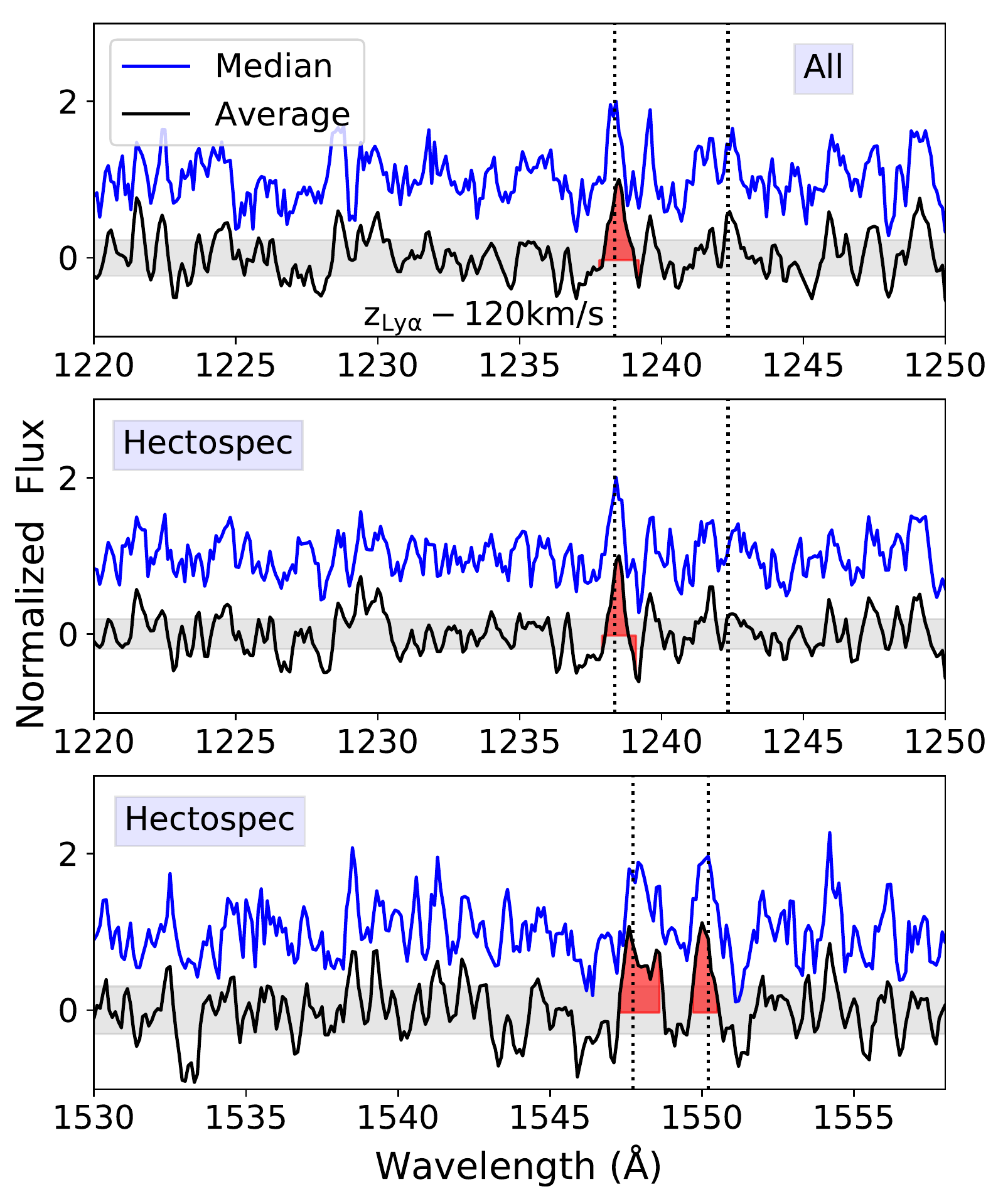}
\caption{
Average and median spectra around \nvld\ (the top and middle panels) and \civl\ (the bottom panel). The top panel shows the composite spectra of all LAEs. The middle and bottom panels show the results of the Hectospec spectra (the M2FS spectra do not cover \civl). All median spectra have been shifted by 1 for clarity. The peak flux of \nvli\ has been normalized to 1. The vertical dotted lines indicate the expected line positions based on the \lya\ redshifts.
\label{fig:NVprofile}
}
\end{figure}

We combine our spectra and search for \nv\ and \civ\ emission lines. Our spectra do not cover the wavelength range for the other lines. The resultant mean and median spectra are shown in Figure~\ref{fig:NVprofile}. In the top panel we plot the spectra at the wavelength range around \nv. The \nvli\ line is detected with S/N $\sim4.6$ in the average spectrum.
The signal is calculated by summing up the pixels around the line within a window of 2 \AA. The noise is estimated from the spectral variation at 1220--1235 \AA\ and 1245--1260 \AA. The flux of \nvli\ is roughly 1\% of the \lya\ flux.
The \nvlii\ line is not detected, as it is usually much weaker than \nvli. We also combine the Hectospec spectra and the M2FS spectra separately, and detect \nvli\ in the two average spectra. The middle panel of Figure~\ref{fig:NVprofile} shows one of them.

In the bottom panel of Figure~\ref{fig:NVprofile}, we plot the average and median spectra around \civ\ based on our Hectospec spectra (the M2FS spectra do not cover this wavelength range). Either of the doublet lines is detected with S/N $\sim3.7$.
The S/N is computed using the same method as we did for \nvli, except that we use a different wavelength range (1530--1545 \AA\ and 1555--1560 \AA) for the noise calculation.
The two lines have a similar flux strength, about 1\% of the \lya\ flux.
LAEs with weaker UV continuum emission tend to have higher \civ\ EWs \citep[e.g.,][]{shiyuba18}. We estimate $M_{1500}$ (absolute magnitude at rest-frame 1500 \AA) for our LAEs based on Figure 7, and find an average $M_{1500}\sim -20$ mag. This is significantly fainter than that in \citet{shiyuba18}, suggesting relatively higher \civ\ EWs in our sample.

As we mentioned earlier, \lya\ is often redshifted compared to systemic redshifts. In Figure~\ref{fig:NVprofile}, we use the vertical dotted lines to denote the expected positions of the lines based on the \lya\ redshifts. We clearly see that both \nvli\ and \civ\ lines are slightly blueshifted relative to \lya. The velocity offset between \nvli\ and \lya\ is $\sim120$ km s$^{-1}$, and the offset between \civ\ and \lya\ is $\sim130$ km s$^{-1}$.
\citet{zheng16} found that the velocity offset between Ly$\alpha$ and \ciii\ is roughly 300 km s$^{-1}$ based on a LAE sample at $z\approx2.8$. The velocity offsets that we found are smaller. This is likely due to the anti-correlation between Ly$\alpha$ EW and velocity offset \citep[e.g.,][]{zhengzheng10,erb14,nakajima18}, because our LAEs have very strong \lya\ emission.

Rest-frame UV emission lines provide powerful constraints of the gas ionization state and metallicity in galaxies \citep[e.g.][]{gutkin16,nakajima18,guo20,mainali20}. The UV lines mentioned earlier usually require hard and intense radiation fields from star formation or AGN.
The line widths of \nv\ and \civ\ in our composite spectra are about $100\sim150$ km s$^{-1}$ (with large uncertainties due to low S/N). There are almost no broad emission lines in the spectra, and thus no detectable AGN broad-line components. In addition, the \civ\ flux compared to \lya\ is much lower than those in typical AGN, including Type 2 AGN. Therefore, AGN contribution (if there is) in our LAE spectra should be small or negligible.
On the other hand, \nv\ and \civ\ are high ionization lines and rarely seen in normal star-forming galaxies. In particular, \nv\ is usually believed to be powered by AGN. For example, \citet{fevre19} combined \ciii\ emitters at $2<z<3.8$ and detected \nv\ in some subsamples. They claimed that their detected \nv\ emission lines are mainly due to narrow-line Type 2 AGN. We cannot rule out a Type 2 AGN contribution in our sample, but this contribution should be small, because of the very low \civ\ flux relative to \lya\ (mentioned above) and the blue UV continuum SEDs of the LAEs.
Nevertheless, currently we are not able to distinguish between the two mechanisms using photoionization models \citep[e.g.,][]{feltre16}, based on only one flux ratio (\nv\ to \civ). More diagnostic lines such as He II and C III] are needed.

It is worth pointing out that the above analysis was based on the composite spectra. We did not detect these UV emission lines in individual LAEs. It is very likely that these lines only exist in a fraction of our LAEs. If so, the above \nv\ and \civ\ flux in the relevant  LAEs would have been largely underestimated.

\section{Summary} \label{sec:conclusions}

We have carried out a spectroscopic survey of LAEs at $z\approx3.1$ in the SXDS field. The LAE candidates were selected by the narrowband technique based on the deep imaging data from Subaru Suprime-Cam. In particular, two narrowband filters NB497 and NB503 were used for target selection. With  spectroscopic observations on MMT Hectospec and Magellan M2FS, we confirmed 150 LAEs. Together with 16 LAEs from \citet{ouchi08}, they form a statistically complete sample of 166 LAEs over a total effective area of $\sim$1.2 deg$^2$. The NB497-band observations cover $\sim$0.5 deg$^2$ and the NB503-band observations cover $\sim$0.7 deg$^2$. This sample is currently the largest spectroscopic confirmed LAE sample at this redshift.

We have constructed a high-quality \lya\ line profile, and calculated \lya\ redshifts by fitting the composite profile to the individual lines. Using the secure redshifts and multi-band photometry, we measured UV slope, \lya\ flux, and EW for each LAE. The \lya\ EW distribution can be described by an exponential form with a scale length of $\sim$63.7 \AA. The median UV slope is $\beta \approx -1.43$.

We have derived a robust \lya\ LF at $z\approx3.1$. We carefully considered four types of sample incompleteness from source detection, candidate selection, spectroscopic observations, and LAE identification. Our LF spans a wide luminosity range from $\sim10^{42.0}$ to $>10^{43.5}$ erg s$^{-1}$ and covers a large area of $\sim$1.2 deg$^2$. The LF can be fit using a Schechter function with $\mathrm{log_{10}(\Phi ^*)=-3.30 ^{+0.09}_{-0.10}}$ and $\mathrm{log_{10}(L_{Ly\alpha}^{*})=42.91 ^{+0.13}_{-0.14}}$, when the faint-end slope $\alpha=-1.6$ is fixed. We have seen significant overdense and underdense regions in our sample, but the wide area coverage of the sample have largely suppressed the effect from such cosmic variance.
Our LF is generally consistent with the results in the literature. At the faint end, it agrees with most previous studies based on narrowband surveys. At the very bright end, our LF is slightly higher than those of many previous studies, showing a density excess compared to the best-fit Schechter function. This excess is likely real and cannot be explained by AGN contribution.

Finally, we stacked the LAE spectra and clearly detected the \nvli\ emission and \civl\ doublet emission lines (S/N $\sim4$). These lines are weak (0.7\%-0.8\% of the \lya\ flux) and narrow ($100\sim150$ km s$^{-1}$). They are rarely seen in normal star-forming galaxies. The detection of these lines in our composite spectra indicate very hard radiation fields in our LAEs on average. More diagnostic lines such as He II and C III] are needed to explore their mechanisms.

\acknowledgments
We acknowledge support from the National Science Foundation of China (11721303, 11890693, 11991052), the National Key R\&D Program of China (2016YFA0400702, 2016YFA0400703), and the Chinese Academy of Sciences (CAS) through a China-Chile Joint Research Fund \#1503 administered by the CAS South America Center for Astronomy in Santiago, Chile. We thank M. Ouchi for providing the transmission curves of the NB497 and NB503 filters. We thank Z. Zheng for helpful discussions. Observations reported here were obtained in part at the MMT Observatory, a joint facility of the University of Arizona and the Smithsonian Institution. This paper includes data gathered with the 6.5 meter Magellan Telescopes located at Las Campanas Observatory, Chile. This research includes data obtained through the Telescope Access Program (TAP), which has been funded by the National Astronomical Observatories of China (the Strategic Priority Research Program `The Emergence of Cosmological Structures', Grant No. XDB09000000), and the Special Fund for Astronomy from the Ministry of Finance.

\facilities{MMT (Hectospec), Magellan:Clay (M2FS)}

\clearpage

\bibliography{ref}{}
\bibliographystyle{aasjournal}

\end{document}